\documentclass[preprint,prd,nofootinbib,showpacs]{revtex4}
\usepackage{graphicx}
\usepackage{epsfig}
\usepackage{bm}
\begin{document}   

\title{\large\bf $B_c \to (J/\Psi,\,\eta_c)\tau\nu$ semileptonic decays within Standard model and beyond.}
\author{Rupak~Dutta${}^{1}$}
\email{rupak@phy.nits.ac.in}
\author{Anupama~Bhol${}^{2}$}
\email{anupama.phy@gmail.com}   
\affiliation{
${}^1$National Institute of Technology Silchar, Silchar 788010, India\\
${}^2$C.~V.~Raman College of Engineering, Bhubaneswar, Odisha 752054, India 
}

\begin{abstract}
Deviations from the standard model prediction have been observed not only in $b \to c$ charged current interactions but also in
$b \to s$ flavor changing neutral current interactions. In particular, the deviation observed in the measured ratio of branching 
fractions $R_D = \mathcal B(B \to D\tau\nu)/\mathcal B(B \to D\,l\,\nu)$ and 
$R_{D^{\ast}} =\mathcal B(B \to D^{\ast}\tau\nu)/\mathcal B(B \to D^{\ast}\,l\,\nu)$, where $l = (e,\,\mu)$, is more pronounced 
and the combined excess currently stands at $3.9\sigma$ level. If it persists and confirmed by future 
experiments, it would be a definite hint of new physics. In this context, we
consider $B_c \to \eta_c\,l\,\nu$ and $B_c \to J/\Psi\,l\,\nu$ decays mediated via $b \to c\,l\,\nu$ charged current interactions 
and employ the most general effective Lagrangian in the presence of new physics to give prediction on 
various observables such as ratio of branching ratio, tau polarization fraction, and forward backward asymmetry for these decay modes.
\end{abstract}
\pacs{%
14.40.Nd, 
13.20.He, 
13.20.-v} 

\maketitle

\section{Introduction}
\label{int}
Although, no direct evidence of new physics has been reported so far, there still exists some discrepancies with the standard model~(SM) 
prediction.
In particular, deviations from the SM expectation in both charged current $b \to c\tau\nu$ transitions as well as neutral current 
$b \to s\,l\bar{l}$ transitions have been observed in various measurements. 
The decays $B \to (D,\,D^{\ast})\tau\nu$ and the lepton flavor universality ratios $R_D$ and 
$R_{D^{\ast}}$ have been studied by BABAR~\cite{Lees:2012xj,Lees:2013uzd}, BELLE~\cite{Huschle:2015rga,Sato:2016svk,Abdesselam:2016xqt}, 
and LHCb~\cite{Aaij:2015yra} experiments. Various measurements of $R_D$ and 
$R_{D^{\ast}}$ are collected in Table.~\ref{tab1}.
\begin{table}[htdp]
\begin{center}
\begin{tabular}{|c|c|c|}
\hline
Experiments & $ R_{D^{\ast}} $ & $R_D$ \\[0.2cm]
\hline
\hline
BABAR &$0.332\pm 0.024\pm 0.018$ &$0.440 \pm 0.058 \pm 0.042 $ \\[0.2cm]
BELLE &$0.293 \pm 0.038 \pm 0.015$ &$0.375 \pm 0.064 \pm 0.026$ \\[0.2cm]
BELLE &$0.302 \pm 0.030 \pm 0.011 $ & \\[0.2cm]
LHCb &$0.336 \pm 0.027 \pm 0.030 $ & \\[0.2cm]
BELLE &$0.276 \pm 0.034^{+ 0.029}_{-0.026}$ & \\[0.2cm]
\hline
\hline
AVERAGE &$0.310 \pm 0.015 \pm 0.008$ &$0.403 \pm 0.040 \pm 0.024 $ \\[0.2cm]
\hline
\end{tabular}
\end{center}
\caption{Current status of $R_D$ and $R_{D^{\ast}}$~\cite{hfag}.}
\label{tab1}
\end{table}
The first unquenched lattice determination of the ratio of branching ratio $R_D = 0.299 \pm 0.011$~\cite{Lattice:2015rga} was reported 
by FNAL/MILC collaboration which is in excellent agreement with the the value of $R_D = 0.300 \pm 0.008$~\cite{Na:2015kha} reported by
HPQCD collaboration. 
In Ref.~\cite{Bigi:2016mdz}, the authors obtain $R_D=0.299 \pm 0.003$ by combining the two lattice calculations, with the experimental 
form factor of the $B \to D\,l\,\nu$ 
from BABAR and BELLE. The result is compatible with the results above, but more accurate.
The FLAG working group combine the two lattice calculations and report the value of $R_D$ to be $0.300 \pm 0.008$~\cite{Aoki:2016frl}.
The SM prediction for $R_{D^{\ast}}$ is $0.252 \pm 0.003$~\cite{Fajfer:2012vx}. At present, the deviation of the measured values of 
$R_D$ and $R_{D^{\ast}}$ from the SM expectation exceeded by $2.2\sigma$ and $3.4\sigma$ respectively~\cite{hfag}. 
Considering the $R_D$-$R_{D^{\ast}}$ 
correlation, the difference with the SM predictions currently stands at about $3.9\sigma$~\cite{hfag}. For theoretical implications of 
these anomalies, we refer to Refs.~\cite{Fajfer:2012vx,Fajfer1, Hou, Akeroyd, tanaka, 
Nierste, miki, Wahab, Deschamps, Blankenburg, Ambrosio, Buras, Pich, Jung, Crivellin, datta, datta1, datta2, datta3, 
datta4, fazio, Crivellin1,  Celis, He, dutta, Tanaka:2016ijq, Deshpand:2016cpw,  Li:2016vvp, Du:2015tda, 
Bernlochner:2015mya, Soffer:2014kxa, Bordone:2016tex, Bardhan:2016uhr, Alok:2016qyh, Ivanov:2015tru, Ivanov:2016qtw, 
Boucenna:2016wpr, Boucenna:2016qad,Nandi:2016wlp,Dutta:2016eml,Alonso:2016oyd,Becirevic:2016yqi,Celis:2016azn} and references therein. 
Very recently, the first measurement of the tau polarization fraction
$P_{\tau}^{D^{\ast}} = -0.44 \pm 0.47^{+0.20}_{-0.17}$ in the decay $B \to D^{\ast}\tau\nu$ was reported by BELLE~\cite{Abdesselam:2016xqt}.

$B_c$ meson, a pseudoscalar ground state composed of two heavy quarks $b$ and $c$, first observed by CDF collaboration in $p\bar{p}$
collisions~\cite{Abe:1998fb}, has a promising prospect on the hadron colliders as 
around $5\times 10^{10}$ $B_c$ events per year are expected at LHC experiments~\cite{Gouz:2002kk,Altarelli:2008xy}. 
Being composed of two heavy quarks, $B_c$ meson
has the unique ability to decay via both $b$ and $c$ quark. Although the $b$ decays are cabbibo suppressed, the charm quark decays, 
however, are cabbibo favored decays as the CKM matrix element $V_{cs} = 1$ is much larger than $V_{cb} = 0.04$. The estimates of the 
$B_c$ total decay width indicate that the $c$ quark transitions provide the dominant contribution while the $b$ quark transitions and 
weak annihilation contribute less. The $c$ quark decays provide around $70\%$ to the total decay width of $B_c$ meson~\cite{Gouz:2002kk}.
Although an indirect constraint can be imposed on various new physics~(NP) from the experimentally measured total decay width of $B_c$ meson, 
however, measurement of various taunic decays of $B_c$ meson in future will give direct access to the beyond the SM physics. The mean 
lifetime of $B_c$ meson $\tau_{B_c} = 0.52^{+0.18}_{-0.12}\,{\rm ps}$ in the SM, calculated using operator product expansion and 
non relativistic QCD~\cite{Bigi:1995fs,Beneke:1996xe,Chang:2000ac}, is consistant with
the measured mean lifetime  $\tau_{B_c} = 0.507(8){\rm ps}$~\cite{pdg}. One can infer from this calculation that no more than $5\%$ 
of the total decay width of $B_c$ meson can be
explained by the semi(taunic) decays of $B_c$ meson. This was confirmed by various other SM caculations as well~\cite{Gershtein:1994jw,
Kiselev:2000pp}. The constraint,
however, can be relaxed upto around $30\%$ depending on the value of the total decay width of $B_c$ meson that is used as input for the
SM calculation of various partonic transitions. 

The $B_c$ meson and its decays have been widely studied in the literature~\cite{Dhir:2008hh,Hernandez:2006gt,Ivanov:2000aj,
Ivanov:2005fd,Ivanov:2006ni,Ebert:2002pp,Ebert:2003cn,Ebert:2003wc,Du:1988ws,Chang:1992pt,Liu:1997hr,AbdElHady:1999xh,Colangelo:1999zn,
Sun:2008wa,Wang:2008xt,Qiao:2012vt,Xiao:2011zz,Liu:2009qa,Cheng:2005um,Sun:2008ew,Wen-Fei:2013uea,Pathak:2013dra,Hsiao:2016pml}. 
The decays $B_c \to (J/\Psi,\,\eta_c)\,l\,\nu$ are  
mediated via $b \to c\,l\,\nu$ transitions and, in principle, NP effects might enter into these decay modes as well. The SM prediction
of these decay modes are already studied by various authors~\cite{Hernandez:2006gt,Ivanov:2000aj,Ivanov:2005fd,Ivanov:2006ni,Ebert:2003cn,
AbdElHady:1999xh,Colangelo:1999zn,Qiao:2012vt,Wen-Fei:2013uea,Pathak:2013dra,Hsiao:2016pml}. Earlier discussions, however, have not 
looked into possible NP effects in these decay modes. In
this study, we wish to study systematically the effect of NP couplings on various observables such as ratio of branching ratios, forward
backward asymmetry, and $\tau$ polarization fraction pertaining to $B_c \to (J/\Psi,\,\eta_c)\,\tau\,\nu$ decays. To analyse the effect of
NP couplings on various observables, we use the most general effective Lagrangian for the $b \to c\,l\,\nu$ decay processes in the 
presence of NP that is valid at the renormalization scale $\mu = m_b$. We use $2\sigma$ constraint coming from the measured values
of the ratio of branching ratios $R_D$ and $R_{D^{\ast}}$ to explore various NP scenarios. Constraint coming from total decay width
of $B_c$ meson is also discussed in details. We, however, do not use the constraint coming from the measured value of $P_{\tau}^{D^{\ast}}$ 
as the uncertainty associated with this observable reported by BELLE is rather large.

Our paper is organised as follows. In section~\ref{ehha}, we introduce the most general effective Lagrangian for the $b \to c\,l\,\nu$ 
transition decays in the presence of NP. The two body $B_c \to \tau\nu$ and three body $B_c \to (J/\Psi,\,\eta_c)\,l\,\nu$ decay branching
ratios are calculated and reported in section~\ref{ehha}. Various observables such as ratio of branching ratios, forward backward
asymmetries, and the $\tau$ polarization are defined. We report our analysis in section~\ref{rd} with a conclusion and summary in 
section~\ref{con}.
 
\section{Effective weak Lagrangian, helicity amplitudes, and observables}
\label{ehha}
\subsection{Effective weak Lagrangian}
We employ the effective field theory approach for the computation of various decay branching fractions in a model independent way.  
The most general effective weak Lagrangian at energy scale $\mu = m_b$ for the $b \to c\,l\,\nu$
transition decays can be expressed as~\cite{Bhattacharya, Cirigliano}
\begin{eqnarray}
\mathcal L_{\rm eff} &=&
-\frac{4\,G_F}{\sqrt{2}}\,V_{c b}\,\Bigg\{(1 + V_L)\,\bar{l}_L\,\gamma_{\mu}\,\nu_L\,\bar{c}_L\,\gamma^{\mu}\,b_L +
V_R\,\bar{l}_L\,\gamma_{\mu}\,\nu_L\,\bar{c}_R\,\gamma^{\mu}\,b_R 
+
\widetilde{V}_L\,\bar{l}_R\,\gamma_{\mu}\,\nu_R\,\bar{c}_L\,\gamma^{\mu}\,b_L \nonumber \\
&&+
\widetilde{V}_R\,\bar{l}_R\,\gamma_{\mu}\,\nu_R\,\bar{c}_R\,\gamma^{\mu}\,b_R +
S_L\,\bar{l}_R\,\nu_L\,\bar{c}_R\,b_L +
S_R\,\bar{l}_R\,\nu_L\,\bar{c}_L\,b_R +
\widetilde{S}_L\,\bar{l}_L\,\nu_R\,\bar{c}_R\,b_L +
\widetilde{S}_R\,\bar{l}_L\,\nu_R\,\bar{c}_L\,b_R \nonumber \\
&&+ 
T_L\,\bar{l}_R\,\sigma_{\mu\nu}\,\nu_L\,\bar{c}_R\,\sigma^{\mu\nu}\,b_L +
\widetilde{T}_L\,\bar{l}_L\,\sigma_{\mu\nu}\,\nu_R\,\bar{c}_L\,\sigma^{\mu\nu}\,b_R\Bigg\} + {\rm h.c.}\,,
\end{eqnarray}
Neglecting the tensor NP couplings and following the same notation as in Ref.~\cite{dutta}, the effective Lagrangian can be
expressed as
\begin{eqnarray}
\label{leff}
\mathcal L_{\rm eff} &=&
-\frac{G_F}{\sqrt{2}}\,V_{c b}\,\Bigg\{G_V\,\bar{l}\,\gamma_{\mu}\,(1 - \gamma_5)\,\nu_l\,\bar{c}\,\gamma^{\mu}\,b -
G_A\,\bar{l}\,\gamma_{\mu}\,(1 - \gamma_5)\,\nu_l\,\bar{c}\,\gamma^{\mu}\,\gamma_5\,b +
G_S\,\bar{l}\,(1 - \gamma_5)\,\nu_l\,\bar{c}\,b \nonumber \\
&& - G_P\,\bar{l}\,(1 - \gamma_5)\,\nu_l\,\bar{c}\,\gamma_5\,b + 
\widetilde{G}_V\,\bar{l}\,\gamma_{\mu}\,(1 + \gamma_5)\,\nu_l\,\bar{c}\,\gamma^{\mu}\,b -
\widetilde{G}_A\,\bar{l}\,\gamma_{\mu}\,(1 + \gamma_5)\,\nu_l\,\bar{c}\,\gamma^{\mu}\,\gamma_5\,b \nonumber \\
&&+
\widetilde{G}_S\,\bar{l}\,(1 + \gamma_5)\,\nu_l\,\bar{c}\,b - \widetilde{G}_P\,\bar{l}\,(1 + \gamma_5)\,\nu_l\,\bar{c}\,
\gamma_5\,b\Bigg\} + {\rm h.c.}\,,
\end{eqnarray}
where
\begin{eqnarray*} 
&&G_V = 1 + V_L + V_R\,,\qquad\qquad
G_A = 1 + V_L - V_R\,, \qquad\qquad
G_S = S_L + S_R\,,\qquad\qquad
G_P = S_L - S_R\, \nonumber \\
&&\widetilde{G}_V = \widetilde{V}_L + \widetilde{V}_R\,,\qquad\qquad
\widetilde{G}_A = \widetilde{V}_L - \widetilde{V}_R\,, \qquad\qquad
\widetilde{G}_S = \widetilde{S}_L + \widetilde{S}_R\,,\qquad\qquad
\widetilde{G}_P = \widetilde{S}_L - \widetilde{S}_R\,.
\end{eqnarray*}
Here $G_F$ is the Fermi coupling constant and $V_{cb}$ is the CKM matrix element. The new vector and scalar NP interactions that
involve left handed neutrinos are denoted by $V_{L,\,R}$ and $S_{L,\,R}$ NP couplings. Similarly for the right handed neutrinos the
NP interactions are denoted by $\widetilde{V}_{L,\,R}$ and $\widetilde{S}_{L,\,R}$ NP couplings, respectively. All these NP couplings
are defined at the renormalization scale $\mu = m_b$.
In the SM, all the NP couplings will be zero leading to $G_{V,A}=1$, $G_{S,P}=0$ and $\widetilde{G}_{V,A,S,P}=0$. 

\subsection{Helicity amplitudes and observables}
We follow Refs.~\cite{Korner,Kadeer} to calculate the various helicity amplitudes for a $B_q$ meson decaying to a pseudoscalar or to
a vector meson along with a charged lepton and an antineutrino in the final state. Again, in order to calculate
the partial decay width of $B_q \to l\nu$ and differential decay rate of three body $B_q \to (P,\,V)l\nu$ decays, we need information
on various nonperturbative hadronic matrix elements which are parameterized in terms of $B_q$ meson decay constants and $B_q \to (P,\,V)$
transition form factors. We refer to Refs.~\cite{dutta,Wen-Fei:2013uea} for a more detailed discussion. 

In the presence of NP, the partial decay width
of $B_q \to l\,\nu$ and differential decay width of three body $B_{q} \to (P,\,V)\,l\,\nu$ decays, where $P(V)$ stands for a 
pseudoscalar(vector) meson, can be expressed as~\cite{dutta}
\begin{eqnarray}
\Gamma(B_q \to l\nu) &=&
\frac{G_F^2\,|V_{cb}|^2}{8\,\pi}\,f_B^2\,m_l^2\,m_{B_q}\,\Big(1 - \frac{m_l^2}{m_{B_q}^2}\Big)^2\,
\Bigg\{\Big[G_A - \frac{m_{B_q}^2}{m_l\,(m_b(\mu) + m_c(\mu))}\,G_P\Big]^2 \nonumber \\
&&+ 
\Big[\widetilde{G}_A - \frac{m_{B_q}^2}{m_l\,(m_b(\mu) + m_c(\mu))}\,\widetilde{G}_P\Big]^2\Bigg\}\,,
\end{eqnarray}
\begin{eqnarray}
\label{pilnu}
\frac{d\Gamma^P}{dq^2} &=&
\frac{8\,N\,|\overrightarrow{p}_P|\,}{3}\Bigg\{\,H_0^2\,\Big(G_V^2 + \widetilde{G}_V^2\Big)\,\Big(1 + \frac{\,m_l^2}{2\,q^2}\Big) \nonumber\\
&& + \frac{3\,m_l^2}{2\,q^2}\,\Big[ \Big(H_t\,G_V + \frac{\sqrt{q^2}}{m_l}\,H_S\,G_S \Big)^2 + \Big(H_t\,\widetilde{G}_V + \frac{\sqrt{q^2}}
{m_l}\,H_S\,\widetilde{G}_S\Big)^2\Big] \Bigg\}\,
\end{eqnarray}
and
\begin{eqnarray}
\label{vlnu1}
\frac{d\Gamma^V}{dq^2} &=&
\frac{8\,N\,|\overrightarrow{p}_V|}{3}\,\Bigg\{ \mathcal{A}_{AV}^2 + \, \frac{ m_l^2}{2\,q^2}\Big[ \mathcal{A}_{AV}^2 + 3\mathcal{A}_{tP}^2 
\Big] 
+
\widetilde{\mathcal{A}}_{AV}^2 + \, \frac{m_l^2}{2\,q^2}\Big[ \widetilde{\mathcal{A}}_{AV}^2 + 3\mathcal{\widetilde{A}}_{tP}^2 \Big] \Bigg\}
\,,
\end{eqnarray}
where
\begin{eqnarray}
&&N = \frac{G_F^2\,|V_{c\, b}|^2\,q^2}{256\,\pi^3\,m_{B_q}^2}\,\Big(1 - \frac{m_l^2}{q^2}\Big)^2\,, \qquad\qquad
H_0 = \frac{2\,m_{B_q}\,|\overrightarrow{p}_P|}{\sqrt{q^2}}\,F_{+}(q^2) \nonumber \\
&&H_t = \frac{m_{B_q}^2 - m_P^2}{\sqrt{q^2}}\,F_0(q^2)\,, \qquad\qquad
H_S=\frac{m_{B_q}^2 - m_P^2}{m_b(\mu) - m_{c}(\mu)}\,F_0(q^2)\,,\nonumber \\
&&\mathcal{A}_{AV}^2 = \mathcal{A}_0^2\,G_A^2 + \mathcal{A}_\parallel^2\,G_A^2 + \mathcal{A}_\perp^2\,G_V^2 \,, \qquad\qquad
\widetilde{\mathcal{A}}_{AV}^2=\mathcal{A}_0^2\,\widetilde{G}_A^2 + \mathcal{A}_\parallel^2\,\widetilde{G}_A^2 + \mathcal{A}_\perp^2\,
\widetilde{G}_V^2\, ,\nonumber\\
&&\mathcal{A}_{tP}=\mathcal{A}_t\,G_A + \frac{\sqrt{q^2}}{m_l}\,\mathcal{A}_P\,G_P \,,\qquad\qquad
\mathcal{\widetilde{A}}_{tP}=\mathcal{A}_t\,\widetilde{G}_A + \frac{\sqrt{q^2}}{m_l}\,\mathcal{A}_P\,\widetilde{G}_P \,.
\end{eqnarray}
and
\begin{eqnarray}
&&\mathcal{A}_0=\frac{1}{2\,m_V\,\sqrt{q^2}}\Big[\Big(\,m_{B_q}^2-m_V^2-q^2\Big)(m_{B_q}+m_V)A_1(q^2)\,-\,\frac{4M_B^2|\vec p_V|^2}{m_{B_q}+m_V}A_2(q^2)
\Big]\,, \nonumber\\
&&\mathcal{A}_\parallel=\frac{2(m_{B_q}+m_V)A_1(q^2)}{\sqrt 2}\,,\qquad\qquad
\mathcal{A}_\perp=-\frac{4m_{B_q}V(q^2)|\vec p_V|}{\sqrt{2}(m_{B_q}+m_V)}\,,\nonumber\\
&&\mathcal{A}_t=\frac{2m_{B_q}|\vec p_V|A_0(q^2)}{\sqrt {q^2}}\,,\qquad\qquad
\mathcal{A}_P=-\frac{2m_{B_q}|\vec p_V|A_0(q^2)}{(m_b(\mu)+m_c(\mu))}\,.
\end{eqnarray}
Here $|\overrightarrow{p}_{P(V)}| = \sqrt{\lambda(m_{B_q}^2,\,m_{P(V)}^2,\,q^2)}/2\,m_{B_q}$ is the three momentum vector of the outgoing meson 
and $\lambda(a,\,b,\,c) = a^2 + b^2 + c^2 - 2\,(a\,b + b\,c + c\,a)$. 

We define several observables such as ratio of branching ratios and tau polarization fraction for various semileptonic $b \to c$ transition
decays. Those are
\begin{eqnarray}
&&R_{M} = \frac{\mathcal B(B_q \to M\tau\nu)}{\mathcal B(B_q \to M\,l\nu)}\,,\qquad\qquad
P_{\tau}^{M} = \frac{\Gamma^{M}(+) - \Gamma^{M}(-)}{\Gamma^{M}(+) + \Gamma^{M}(-)}\,, 
\end{eqnarray}
where, $l$ is either an electron or a muon and $B_q$ is either a $B$ meson or a $B_c$ meson. Similarly, $M$ refers to the outgoing 
pseudoscalar or vector meson. Again, 
$\Gamma(+)$ and $\Gamma(-)$ denote the decay widths of 
positive and negative helicity $\tau$ lepton, respectively. It is 
also worth mentioning that, for $B_q \to P\tau\nu$ decays, the tau polarization fraction does not depend on $V_{L,\,R}$ and
$\widetilde{V}_{L,\,R}$ NP couplings if we assume that NP effect is coming from new vector interactions only. 
We also construct various $q^2$ dependent observables such as differential branching fractions DBR$(q^2)$, the ratio of branching 
fractions $ R(q^2)$, and the forward-backward asymmetry parameter $A^{FB}(q^2)$ for the $B_{c} \to (\eta_c,\, J/\Psi)\tau\nu$ decays
such that
\begin{eqnarray}
&& DBR(q^2) = \Big(\frac{d\Gamma}{dq^2}\Big)/\Gamma_{tot} \,, \qquad\qquad
R(q^2)=\frac{DBR(q^2)\Big(B \to (P,\,V)\,\tau\,\nu\Big)}{DBR(q^2)\Big(B \to (P,\,V)\,l\,\nu\Big)} \nonumber\\
&&[A^{FB}]_{(P,\,V)}(q^2)=\frac{\Big(\int_{-1}^{0}-\int_{0}^{1}\Big)d\cos \theta_l\frac{d\Gamma^{(P,\,V)}}{dq^2\,d\cos\theta_l}}{\frac{d\Gamma^{(P,\,V)}}{dq^2}}\,.
\end{eqnarray}
In the presence of various NP couplings, the forward backward asymmetry parameter for $B_q \to P\,l\,\nu$ decays can be written as
\begin{eqnarray}
\label{eq:afbplnu}
A^{FB}_P(q^2) &=& \frac{3\,m_l^2}{2\,q^2}\frac{H_0\,G_V\,\Big[\Big(H_t\,G_V + \frac{\sqrt{q^2}}{m_l}\,H_S\,G_S \Big) + \Big(H_t\,
\widetilde{G}_V + \frac{\sqrt{q^2}}{m_l}\,
H_S\,\widetilde{G}_S\Big)\,\Big]}{H_0^2\,(G_V^2+\widetilde{G}_V^2)(1+\frac{m_l^2}{2\,q^2})+\frac{3\,m_l^2}{2\,q^2}\,\Big[\Big(H_t\,G_V + 
\frac{\sqrt{q^2}}{m_l}\,H_S\,G_S \Big)^2 + \Big(H_t\,\widetilde{G}_V + \frac{\sqrt{q^2}}{m_l}\,H_S\,\widetilde{G}_S\Big)^2\,\Big]}\,.
\nonumber \\
\end{eqnarray}
Similarly, for $B_q \to V\,l\,\nu$ decay mode, the explicit expression for the forward backward asymmetry parameter is
\begin{eqnarray}
\label{eq:afbvlnu}
A^{FB}_V(q^2) &=& 
\frac{3}{2}\frac{\mathcal{A}_\parallel\,\mathcal{A}_\perp\,\Big(G_A\,G_V-\widetilde{G}_A\widetilde{G}_V\Big)+\frac{m_l^2}{q^2}\mathcal{A}_0\,
G_A\Big[\mathcal{A}_t\,G_A - \frac{\sqrt{q^2}}{m_l}\,\mathcal{A}_P\,G_P + \mathcal{A}_t\,\widetilde{G}_A - \frac{\sqrt{q^2}}{m_l}\,
\mathcal{A}_P\,\widetilde{G}_P \Big]} 
{\mathcal{A}_{AV}^2 +  \frac{m_l^2}{2\,q^2}\Big[ \mathcal{A}_{AV}^2 + 3\mathcal{A}_{tP}^2 \Big] 
+ \widetilde{\mathcal{A}}_{AV}^2 + \frac{m_l^2}{2\,q^2}\Big[\widetilde{\mathcal{A}}_{AV}^2 + 3\mathcal{\widetilde{A}}_{tP}^2 \Big]}\,
\nonumber \\
\end{eqnarray}
It is worth mentioning that, although, the forward backward asymmetry parameter does depend on all the NP couplings for $B_q \to V\,\tau\nu$ 
decays, it, however, does not depend on $V_{L,\,R}$ and $\widetilde{V}_{L,\,R}$ NP couplings for the $B_q \to P\,\tau\nu$ decays if we assume
that only vector type NP couplings contribute to these decay modes. The 
dependancy gets cancelled in the ratio. The tau polarization fraction and the forward backward asymmetry parameter can, in principle, 
provide useful information regarding the various Lorentz structures of beyond the SM physics. We now proceed to discuss the results of
our analysis.

\section{Numerical calculations}
\label{rd}
We first report in Table.~\ref{tab_in} all the relevant input parameters that are used for our numerical estimates. 
For the quark, lepton, and meson masses, we use 
the most recent values reported in Ref.~\cite{pdg}. Similarly, for the mean lifetime of $B^-$ and $B_c$ meson, we use the values 
reported in Ref.~\cite{pdg}. We use Ref.~\cite{Alonso:2016oyd} for the $B_c$ meson decay constant. The mass and decay constant reported in 
Table.~\ref{tab_in} are in $\rm GeV$ units, whereas, the mean lifetime of $B^-$ and $B_c$ meson are in seconds. The uncertainty 
associated with $f_{B_c}$ and $V_{cb}$ are indicated by the number in parentheses. The errors in all the other input parameters 
are unimportant for us and hence not included in the Table.~\ref{tab_in}.

\begin{table}[htdp]
\begin{center}
\begin{tabular}{||c|c||c|c||c|c||}
\hline
\hline
$m_b(m_b)$ &$4.18 $ &$m_{J/\Psi}$ &$3.0969$ &$m_{{D^{\ast}}^0}$ &$2.00685$ \\[0.2cm]
\hline
$m_c(m_b)$ &$0.91$ &$m_{\eta_c}$ &$2.9834$ &$\tau_{B^-}$ &$1.638 \times 10^{-12}$ \\[0.2cm]
\hline
$m_e$ & $0.510998928\times 10^{-3}$ &$m_{B^-}$ &$5.27931$ &$\tau_{B_c}$ &$0.507 \times 10^{-12}$ \\[0.2cm]
\hline
$m_{\mu}$ &$0.1056583715$ &$m_{B_c}$ &$6.2751$ &$f_{B_c}$ &$0.434(0.015)$ \\[0.2cm]
\hline
$m_{\tau}$ &$1.77682$ &$m_{D^0}$ &$1.86483$ &$V_{cb}$ & $0.0409(0.0011)$\\[0.2cm]
\hline
\end{tabular}
\end{center}
\caption{Theory input parameters}
\label{tab_in}
\end{table}

For the $B_c\to \eta_c$ and $B_c \to J/\Psi$ hadronic form factors, we follow Ref.~\cite{Wen-Fei:2013uea}.
The relevant formula for $F_0(q^2)$, $F_+(q^2)$, $V(q^2)$, $A_0(q^2)$, $A_1(q^2)$, and $A_2(q^2)$ pertinent for our
discussion, taken from Ref.~\cite{Wen-Fei:2013uea} is
\begin{eqnarray}
F(q^2) = F(0)\,\exp\Big[a\,q^2 + b\,(q^2)^2\Big]\,,
\end{eqnarray}
where $F$ stands for the form factors $F_0$, $F_+$, $V$, $A_0$, $A_1$, and $A_2$ and $a$, $b$ are                              
the fitted parameters.
The numerical values of $B_c \to \eta_c$ and $B_c \to J/\Psi$ form factors at $q^2 = 0$ and their fitted parameters $a$ and $b$, 
calculated in perturbative QCD~(PQCD) approach, collected from Ref.~\cite{Wen-Fei:2013uea}, are 
listed in Table~\ref{tab2}. For our numerical analysis, we added the errors in quadrature. We also report the most important 
experimental input parameters $R_D$ and $R_{D^{\ast}}$ with their uncertainties measured by BABAR, BELLE, and LHCb 
in Table.~\ref{tab1}. We use the average values of $R_D$ and $R_{D^{\ast}}$ for our analysis. In our analysis, we added the statistical
and systematic uncertainties in quadrature.
\begin{table}[htdp]
\begin{center}
\begin{tabular}{|c|c|c|c||c|c|c|c|}
\hline
Form factors & $F_0$ & $a$ & $b$&Form factors &$F_0$ &$a$ &$b$\\[0.2cm]
\hline
\hline
$F_0^{B_c \to \eta_c}$ &$0.48\pm 0.06\pm 0.01$ &$0.037 $&$0.0007$&$A_0^{B_c \to J/\Psi}$ &$0.52 \pm 0.02 \pm 0.01$ &$0.047 $&$0.0017$ 
\\[0.2cm]
\hline
$F_+^{B_c \to \eta_c}$ &$0.48 \pm 0.06 \pm 0.01$ & $0.055$&$0.0014$& $A_1^{B_c \to J/\Psi}$ &$0.46 \pm 0.02 \pm 0.01$ &$0.038 $&$0.0015$ 
\\[0.2cm]
\hline
$V^{B_c \to J/\Psi}$ &$0.42 \pm 0.01 \pm 0.01$ &$0.065 $&$0.0015$& $A_2^{B_c \to J/\Psi}$ &$0.64 \pm 0.02 \pm 0.01$ &$0.064 $&$0.0041$ 
\\[0.2cm]
\hline
\hline
\end{tabular}
\end{center}
\caption{$B_c \to \eta_c$ and $B_c \to J/\Psi$ form factors at $q^2 = 0$ taken from Ref.~\cite{Wen-Fei:2013uea}.}
\label{tab2}
\end{table}

The SM branching ratios, ratio of branching ratios, and the tau polarization fraction for all the relevant decay modes are presented
in Table.~\ref{tab3}. Uncertainties in each observable may come from mainly two different sourses: first it may come from not very well
known input parameters such as CKM matrix elements and second it may come from the hadronic input parameters such as meson to meson 
form factors and meson decay constants. To see the effect of above mentioned uncertainties on various observables, we perform a random
scan of all the input parameters such as CKM matrix element, form factors, and decay constants within $1\sigma$ of their central values.
The central values of all the observables obtained using the central values of all the input parameters and the $1\sigma$ range obtained
from our random scan are reported in Table.~\ref{tab3}.
\begin{table}[htdp]
\begin{center}
\begin{tabular}{|c|c|c||c|c|c|}
\hline
Observables & Central value &$1\sigma$ range &Observables &Central value &$1\sigma$ range\\[0.2cm]
\hline
\hline
$\mathcal B(B_c \to \tau\nu)\times 10^2$ &$2.20$ &$[1.95, 2.48]$ & $R_{\eta_c}$ &$0.308$ &$[0.235, 0.429]$\\[0.2cm]
\hline
$\mathcal B(B_c \to \eta_c\,l\nu)\times 10^3$ &$4.85$ &$[3.50, 6.49]$ &$R_{J/\Psi}$ &$0.289$ &$[0.279, 0.301]$\\[0.2cm]
\hline
$\mathcal B(B_c \to \eta_c\,\tau\nu)\times 10^3$ &$1.49$ &$[1.09, 1.99]$ &$P_{\tau}^{\eta_c}$ &$0.345$ &$[0.141, 0.530]$\\[0.2cm]
\hline
$\mathcal B(B_c \to J/\Psi\,l\nu)\times 10^3$ &$11.36$ &$[9.44, 13.53]$ &$P_{\tau}^{J/\Psi}$ &$-0.465$ &$[-0.433, -0.492]$\\[0.2cm]
\hline
$\mathcal B(B_c \to J/\Psi\,\tau\nu)\times 10^3$ &$3.29$ &$[2.80, 3.83]$ &$P_{\tau}^D$ &$0.336$ & $[0.334, 0.338]$\\[0.2cm]
\hline
 & & &$P_{\tau}^{D^{\ast}}$ &$-0.505$ & $[-0.475, -0.532]$\\[0.2cm]
\hline
\hline
\end{tabular}
\end{center}
\caption{SM prediction of various observables}
\label{tab3}
\end{table}

We wish to determine the NP effect on each observable in a model independent way. We assume four different NP 
scenarios. All the NP couplings are assumed to be real for our analysis. Again, we consider that NP affects the third 
generation leptons only. The allowed NP parameter space is obtained by imposing $2\sigma$ constraint coming from
the measured values of the ratio of branching ratios $R_D$ and $R_{D^{\ast}}$. This automatically guarantee that the
resulting NP parameter space can simultaneously explain the anomalies persisted in $R_D$ and $R_{D^{\ast}}$. Now we
proceed to discuss various NP scenarios.

\subsection{Scenario I: only $V_L$ and $V_R$ type NP couplings}
\label{vl_vr}
\begin{figure}
\begin{center}
\includegraphics[width=6cm,height=5cm]{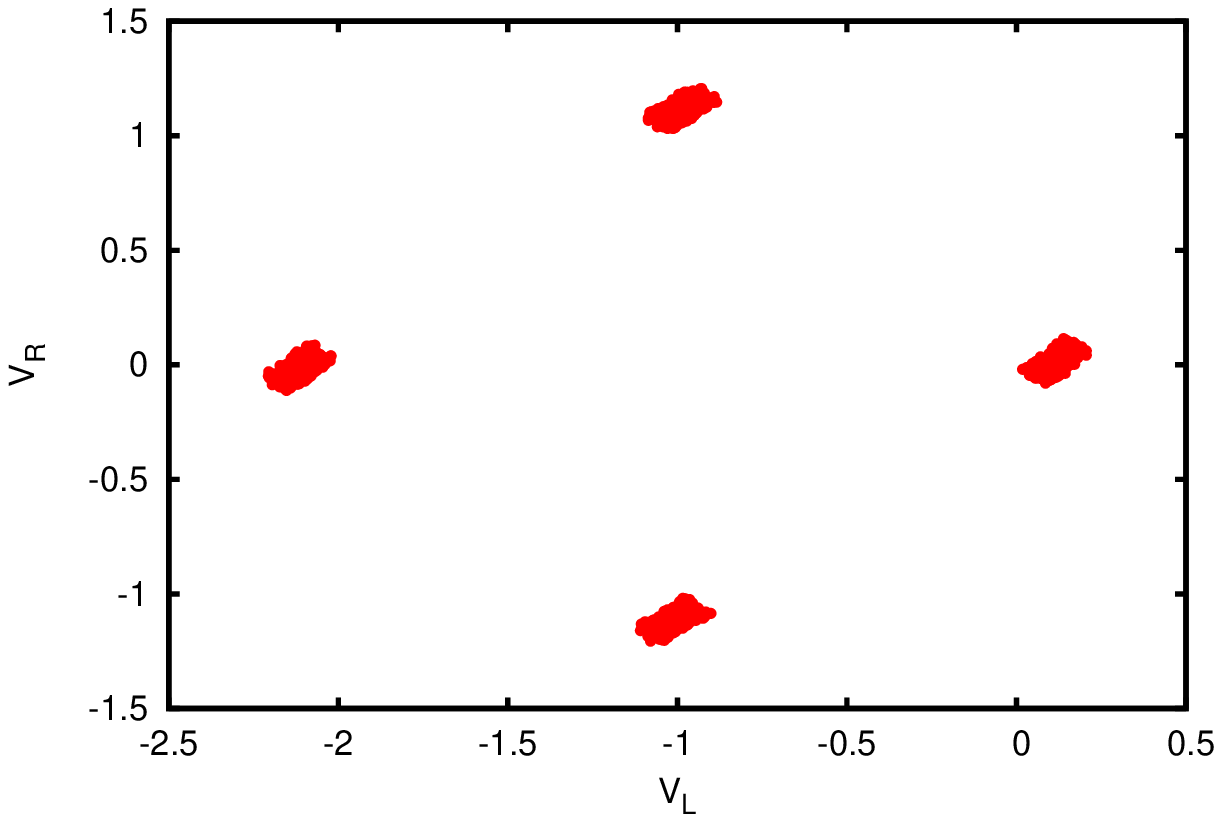}
\includegraphics[width=6cm,height=5cm]{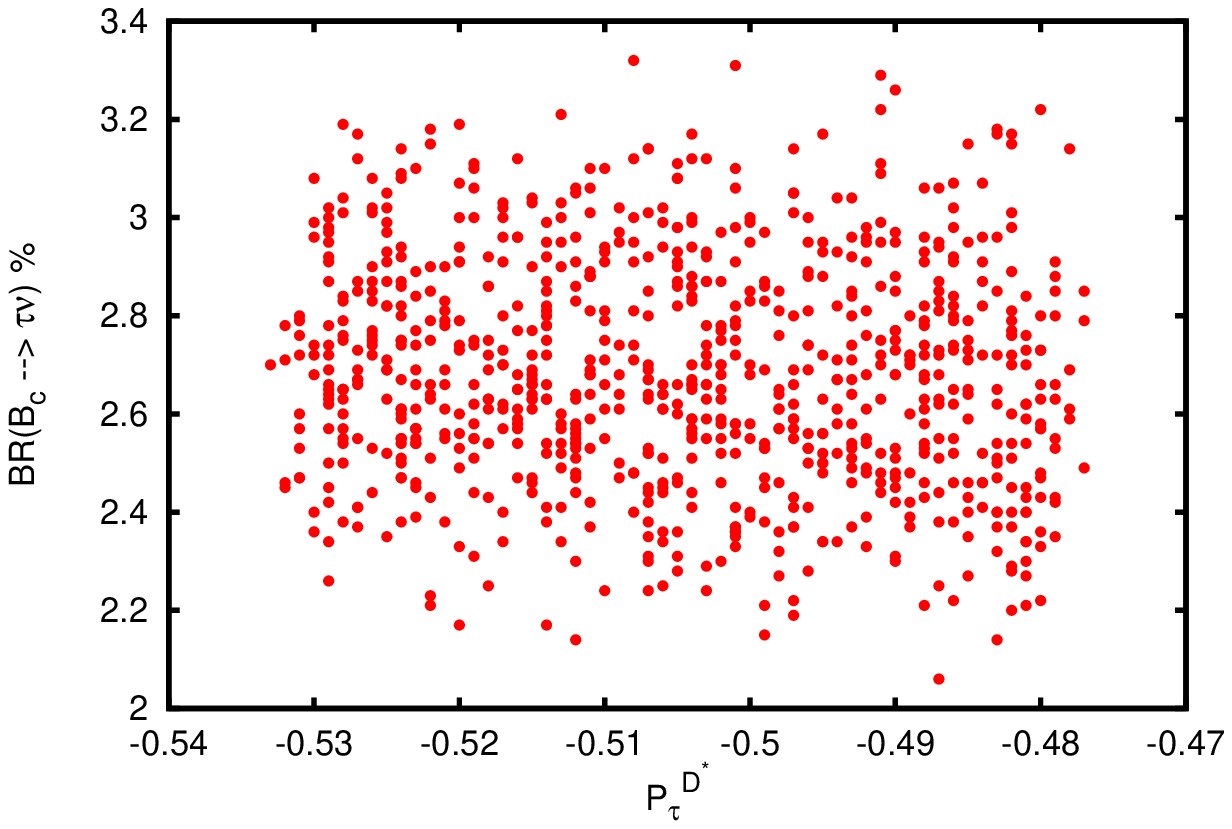}
\end{center}
\caption{Allowed ranges of $V_L$ and $V_R$ NP couplings are shown in the left panel once $2\sigma$ constraint coming from the measured
values of the ratio of branching ratios $R_D$ and
$R_{D^{\ast}}$ is imposed. We show in the right panel the allowed ranges in $\mathcal B(B_c \to \tau\nu)$ and $P_{\tau}^{D^{\ast}}$ in
the presence of these NP couplings.}
\label{vlvr}
\end{figure}

In this scenario, we have considered the effect of only $V_L$ and $V_R$ type NP couplings on various observables. In the left panel of
Fig.~\ref{vlvr}, we
show the allowed range of new vector couplings $V_L$ and $V_R$ that satisfies the $2\sigma$ experimental constraint coming from 
$R_D$ and $R_{D^{\ast}}$.
The range of each observable for $V_L$ and $V_R$ type NP couplings is tabulated in Table.~\ref{tab4}. We also show in the right panel of
Fig.~\ref{vlvr} the allowed ranges of $\mathcal B(B_c \to \tau\nu)$ and the tau polarization fraction $P_{\tau}^{D^{\ast}}$. We want to 
emphasize 
that the central value of $P_{\tau}^{D^{\ast}}$ reported by BELLE lies outside the allowed range of $P_{\tau}^{D^{\ast}}$ obtained in 
this scenario. However, the measured $1\sigma$ range of the observable $P_{\tau}^{D^{\ast}}$ does overlap with the allowed range. Again, 
the uncertainty associated with the measured value of $P_{\tau}^{D^{\ast}}$ is rather large.
The allowed range of $\mathcal B(B_c \to \tau\nu)$ is also compatible with the total decay width of $B_c$ meson.
As expected, the tau polarization fraction pertaining to $B \to D\tau\nu$ and $B_c \to \eta_c\tau\nu$ decays does not vary at all 
as the NP effects coming from $V_L$ and $V_R$ couplings cancel in the ratios.

\begin{table}[htdp]
\begin{center}
\begin{tabular}{|c|c||c|c||c|c|}
\hline
Observables &  Range &Observables & Range &Observables & Range\\[0.2cm]
\hline
\hline
$\mathcal B(B_c \to \tau\nu)\times 10^2$ &$[2.06, 3.32]$ &$R_{\eta_c}$ & $[0.240, 0.658]$& $P_{\tau}^{J/\Psi}$ &$[-0.435, -0.491]$\\[0.2cm]
\hline
$\mathcal B(B_c \to \eta_c\,\tau\nu)\times 10^3$ &$[1.14, 2.97]$ &$R_{J/\Psi}$ &$[0.300, 0.413]$ & $P_{\tau}^D$ &$[0.334, 0.338]$ \\[0.2cm]
\hline
$\mathcal B(B_c \to J/\Psi\,\tau\nu)\times 10^3$ &$[3.12, 5.09]$ &$P_{\tau}^{\eta_c}$ &$[0.141, 0.530]$ &$P_{\tau}^{D^{\ast}}$ & 
$[-0.477, -0.533]$\\[0.2cm]
\hline
\hline
\end{tabular}
\end{center}
\caption{Allowed ranges of various observables in the presence of $V_L$ and $V_R$ NP couplings}
\label{tab4}
\end{table}

\begin{figure}
\begin{center}
\includegraphics[width=5cm,height=4cm]{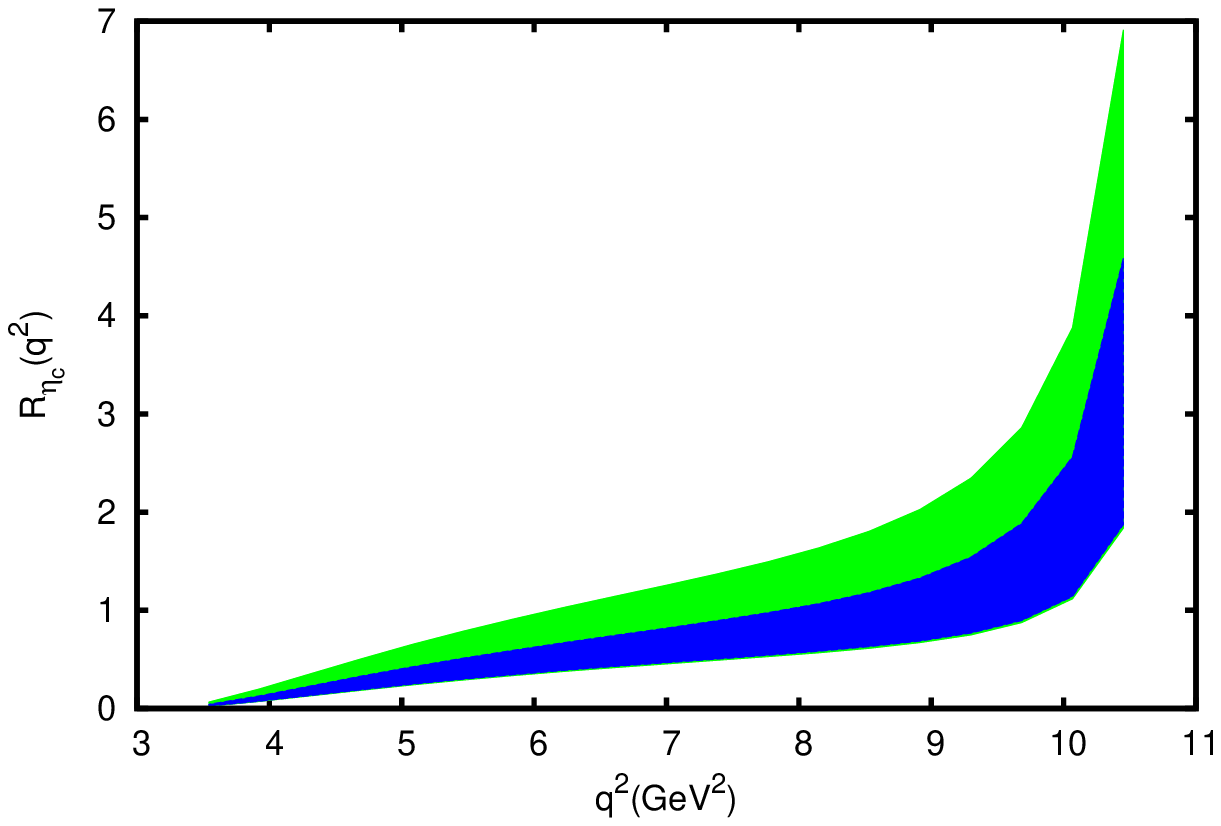}
\includegraphics[width=5cm,height=4cm]{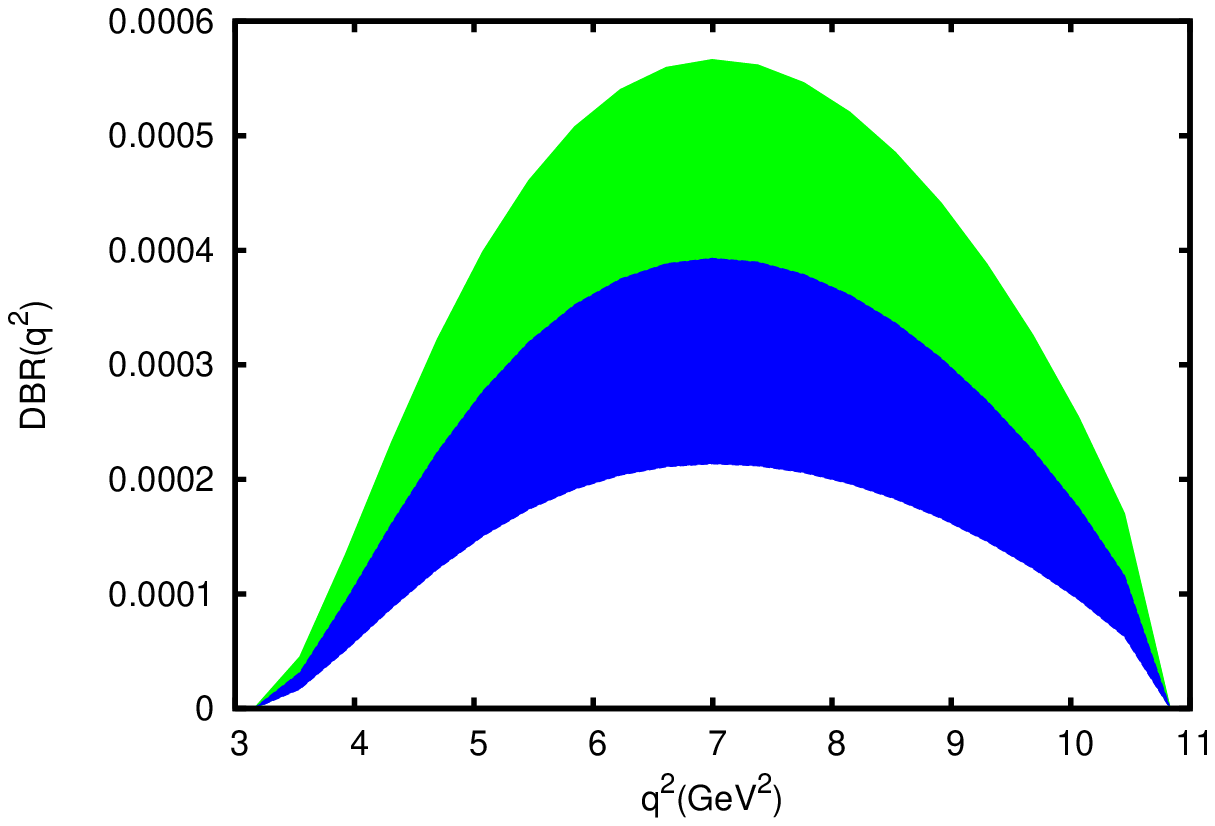}
\includegraphics[width=5cm,height=4cm]{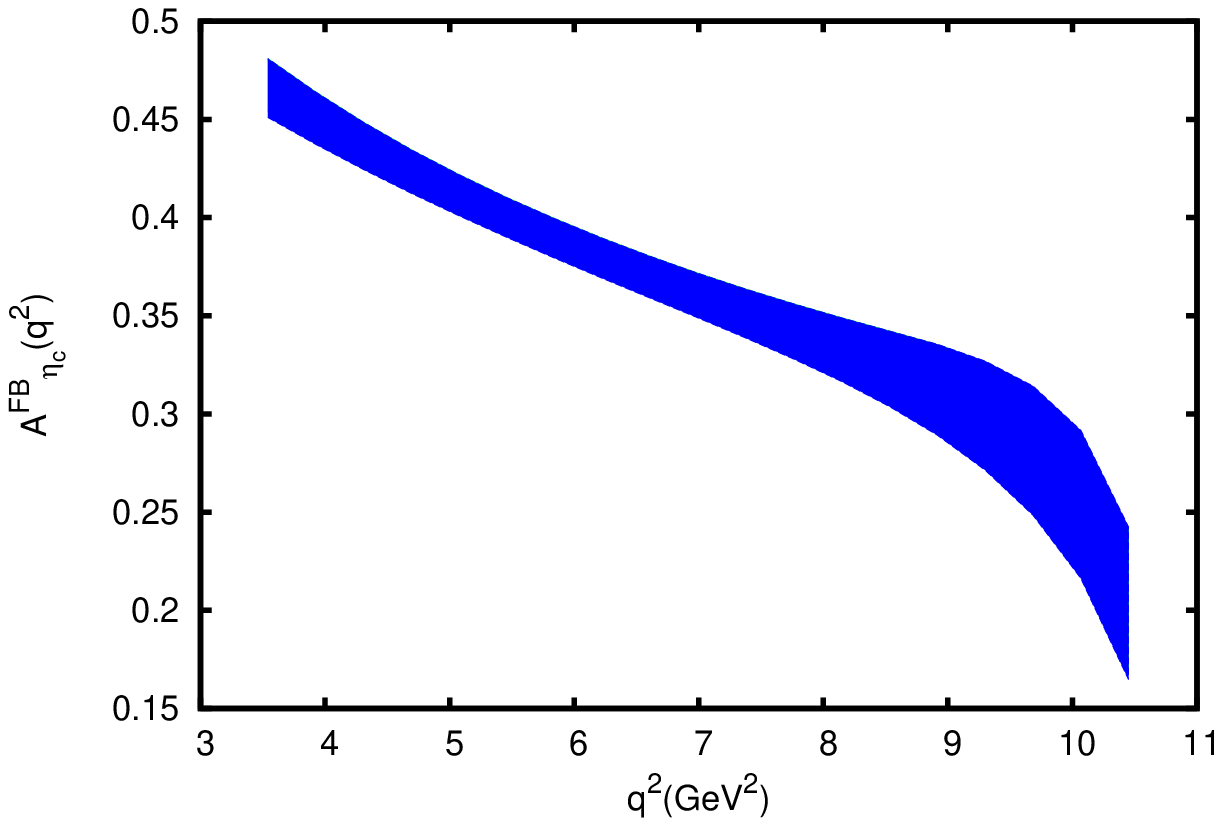}
\includegraphics[width=5cm,height=4cm]{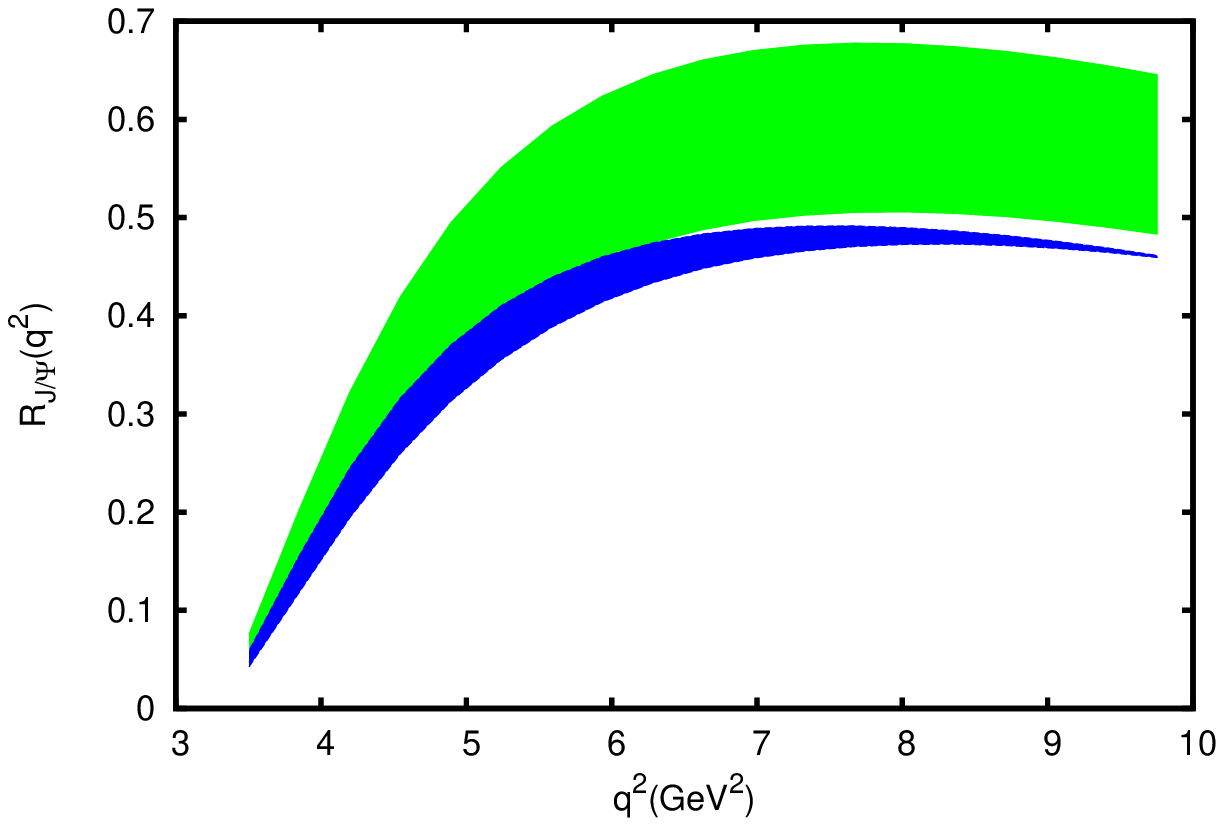}
\includegraphics[width=5cm,height=4cm]{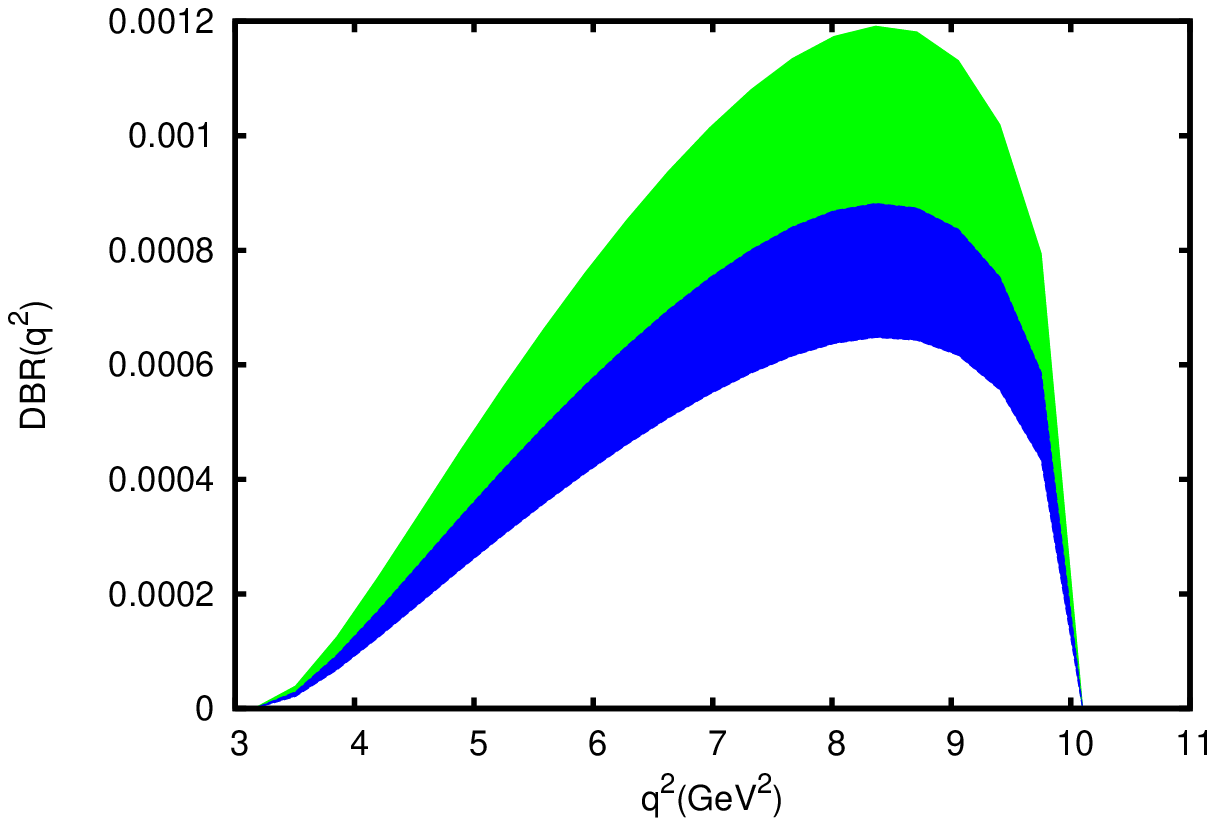}
\includegraphics[width=5cm,height=4cm]{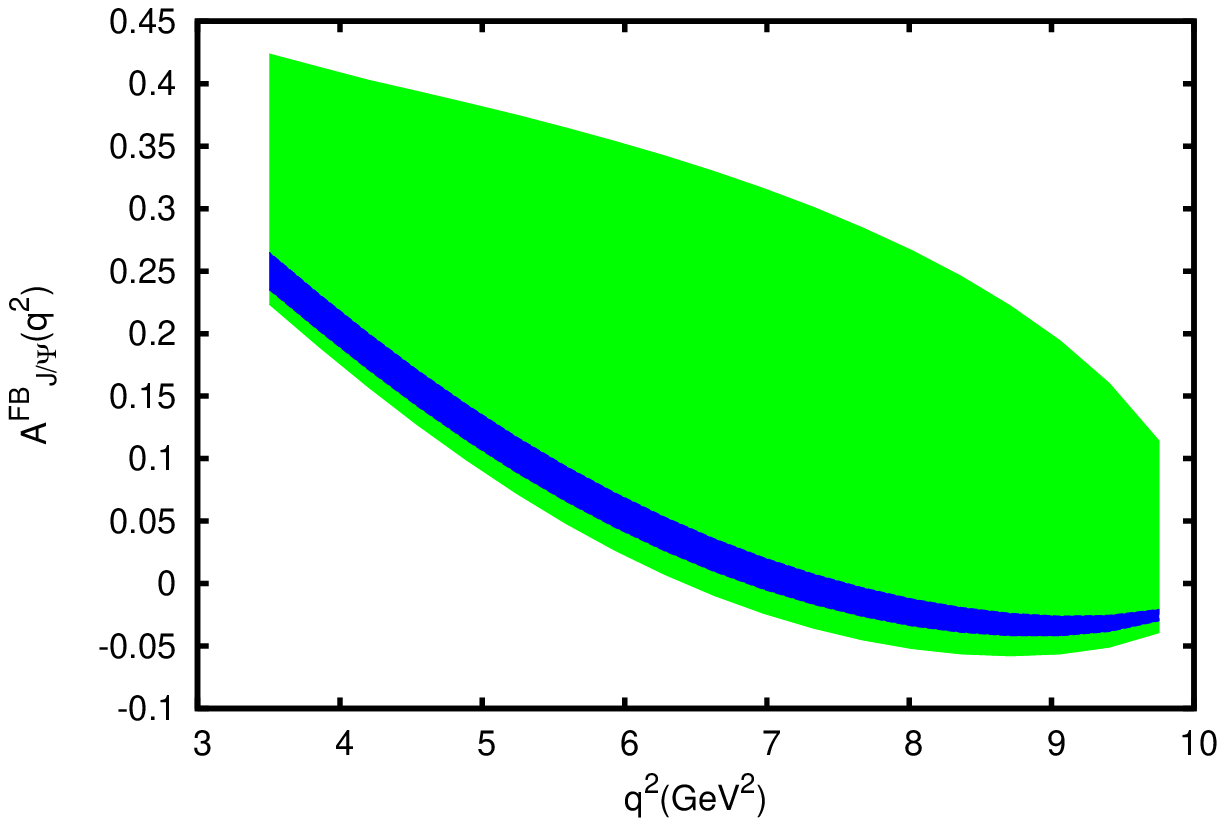}
\end{center}
\caption{Range in various $q^2$ dependent observables such as DBR$(q^2)$, $R(q^2)$, and $A^{FB}(q^2)$ for the $B_c \to \eta_c\tau\nu$~
(upper panel) and $B_c \to J/\Psi\tau\nu$~(lower panel) decays. The allowed range is each observable is shown in light~(green)
band once the NP couplings $(V_L,\,V_R)$ are varied within the allowed ranges shown in the left panel of Fig.~\ref{vlvr}. We show
in dark~(blue) band the corresponding SM prediction.}
\label{obs_vlvr}
\end{figure}

In Fig.~\ref{obs_vlvr}, we show the effect of $V_L$ and $V_R$ NP couplings on various observables such as ratio of branching
ratio $R(q^2)$, the forward backward asymmetry $A^{FB}(q^2)$, and differential branching ratio ${\rm DBR}(q^2)$ as a function of $q^2$
for the $B_c \to \eta_c\tau\nu$ and $B_c \to J/\Psi\tau\nu$ decays. We show in dark~(blue) band the SM range and show in light~(green) 
band the 
allowed range of each observable once the NP couplings $V_L$ and $V_R$ are switched on. We see significant deviation from the 
SM prediction of all the 
observables. The forward backward asymmetry parameter, $A^{FB}(q^2)$, does not vary with the NP couplings $V_L$ and $V_R$ for the 
$B_c \to \eta_c\tau\nu$ decay mode. It is expected as the NP dependency cancels in the ratio since $B_c \to \eta_c\tau\nu$ 
decay mode depends on $G_V$ couplings only.  

\subsection{Scenario II: only $S_L$ and $S_R$ type NP couplings}
\label{sl_sr}
\begin{figure}
\begin{center}
\includegraphics[width=6cm,height=5cm]{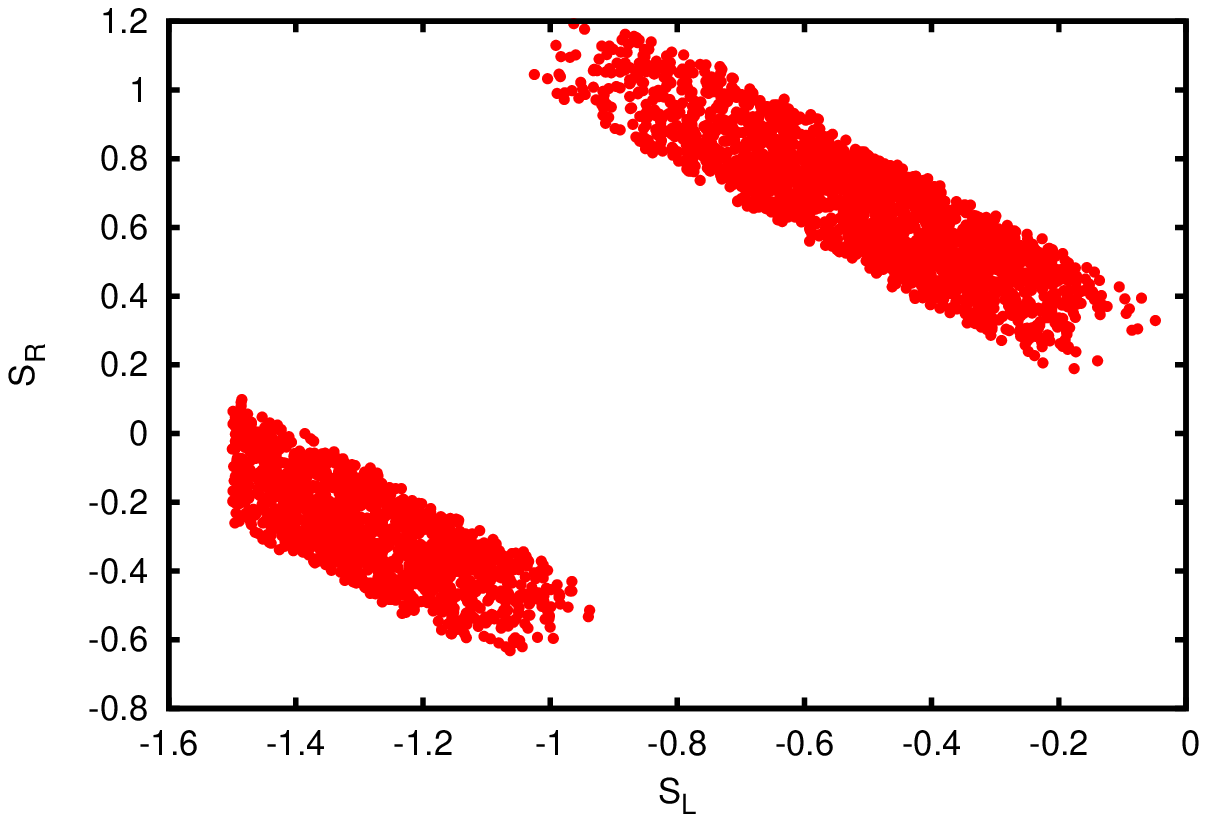}
\includegraphics[width=6cm,height=5cm]{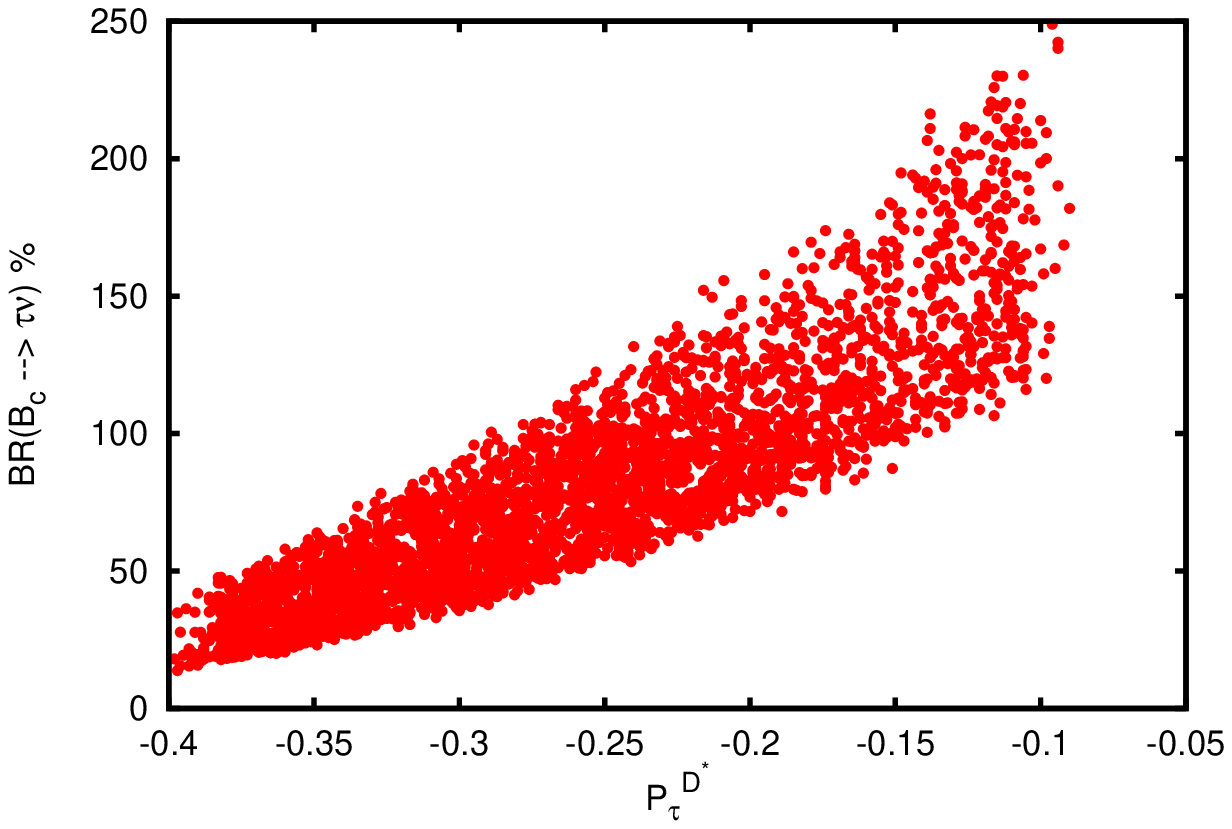}
\end{center}
\caption{Allowed ranges of $S_L$ and $S_R$ NP couplings are shown in the left panel once $2\sigma$ constraint coming from the measured
values of the ratio of branching ratios $R_D$ and
$R_{D^{\ast}}$ is imposed. We show in the right panel the allowed ranges in $\mathcal B(B_c \to \tau\nu)$ and $P_{\tau}^{D^{\ast}}$ in
the presence of these NP couplings.}
\label{slsr}
\end{figure}

In this scenario, we vary only the new scalar interactions $S_L$ and $S_R$ while keeping all other NP couplings to be zero. We restrict
the $S_L$ and $S_R$ parameter space using the $2\sigma$ experimental constraint coming from measured values of $R_D$ and $R_{D^{\ast}}$.
The allowed range of $S_L$ and $S_R$ is shown in the left panel of Fig.~\ref{slsr}. We also show in the right panel of Fig.~\ref{slsr}
the allowed ranges of $\mathcal B(B_c \to \tau\nu)$ and $P_{\tau}^{D^{\ast}}$ in this scenario. We see significant deviation of all the
observables from the SM expectation in this scenario. It is also worth mentioning that the tau
polarization $P_{\tau}^{D^{\ast}}$ deviates significantly from the central value reported by BELLE. However, the uncertainty associated
with the measured value of $P_{\tau}^{D^{\ast}}$ is rather large. Again, we notice that, in this scenario, the value of 
$\mathcal B(B_c \to \tau\nu)$ can exceed the total decay width of $B_c$ meson for some particular values of $S_L$ and $S_R$. We note that
only $\le 5\%$ of the total decay width of $B_c$ meson can be explained by semitaunic decays. However, this constraint can be relaxed
upto $30\%$. If we assume that $\mathcal B(B_c \to \tau\nu)$ can not 
be greater than $5\%$, then although $S_L$ and $S_R$ type NP couplings can explain the anomalies in $R_D$ and 
$R_{D^{\ast}}$, it, however, can not accommodate $\mathcal B(B_c \to \tau\nu)$. Even with $30\%$ constraint, a large part of the NP parameter 
space prefered by $R_D$ and $R_{D^{\ast}}$ can be excluded. The allowed ranges of each observable obtained 
in the presence of $S_L$ and $S_R$ NP couplings are tabulated in Table.~\ref{tab5}.
 
\begin{figure}
\begin{center}
\includegraphics[width=5cm,height=4cm]{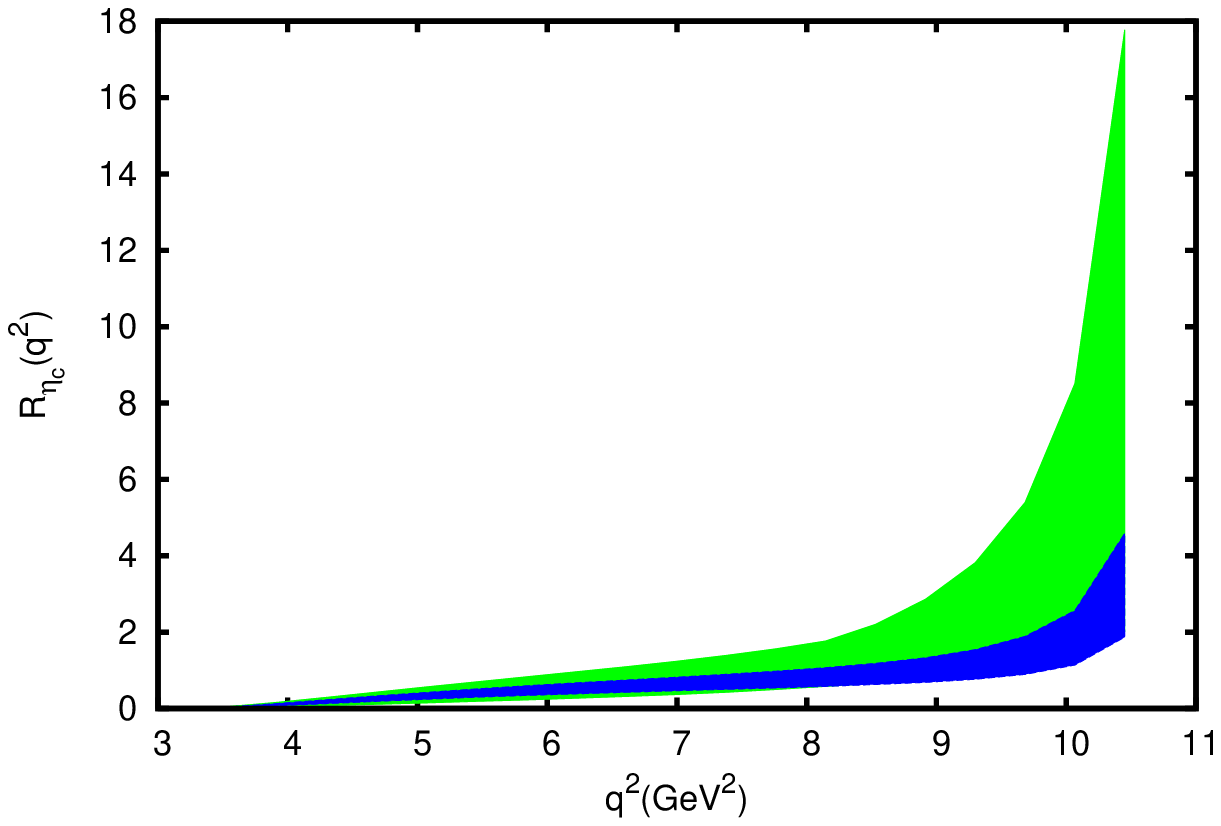}
\includegraphics[width=5cm,height=4cm]{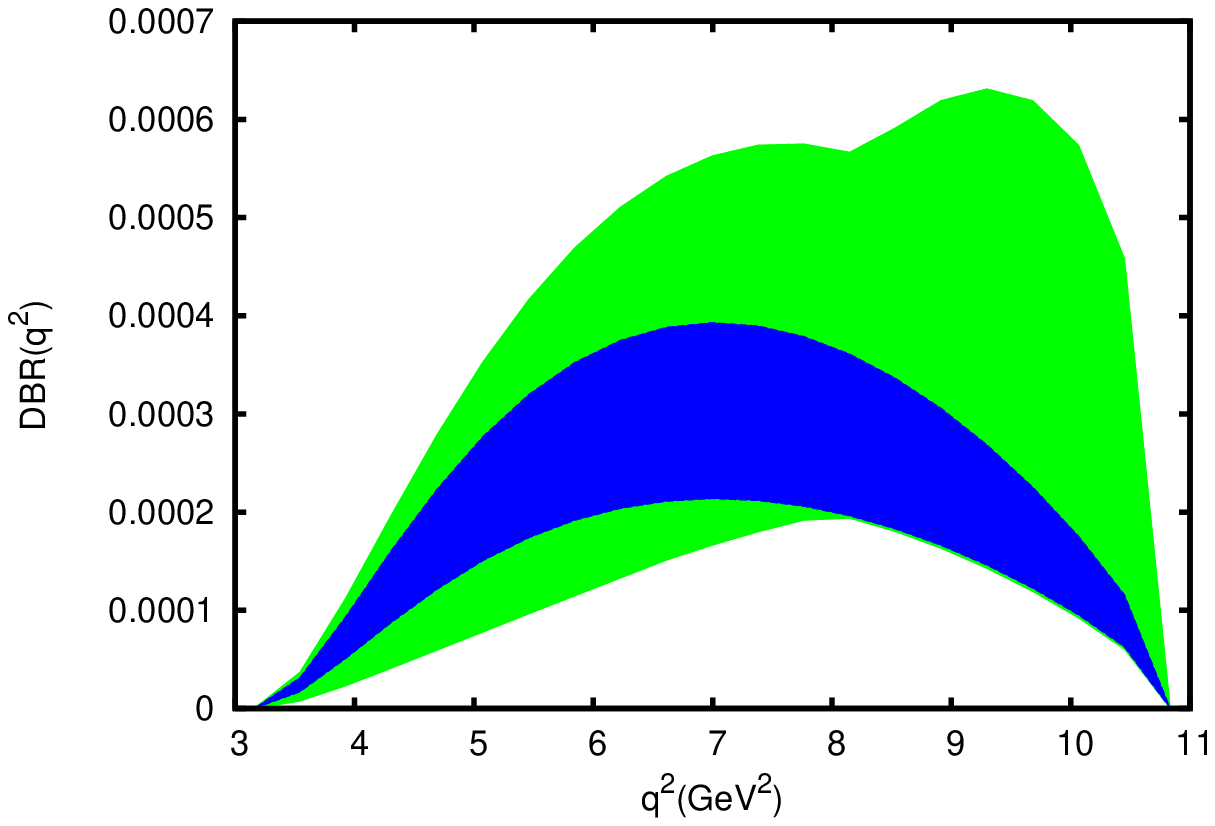}
\includegraphics[width=5cm,height=4cm]{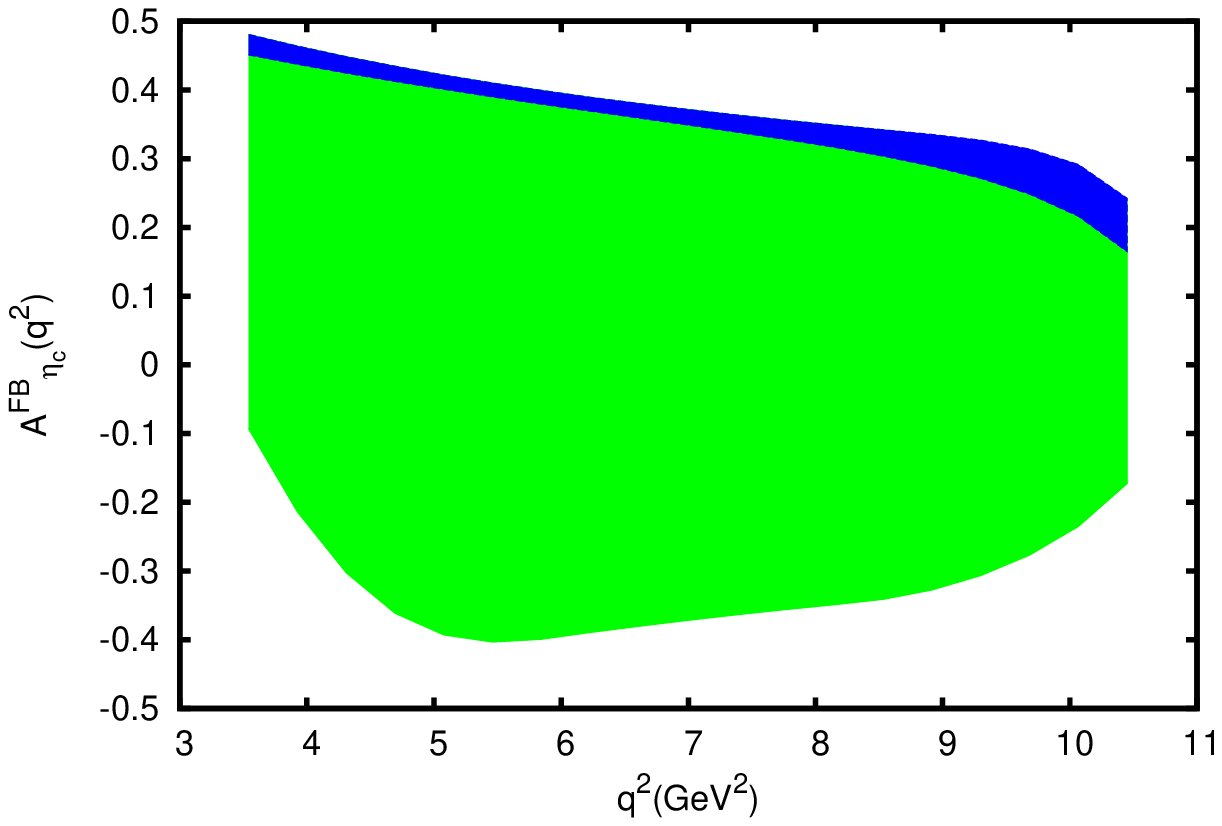}
\includegraphics[width=5cm,height=4cm]{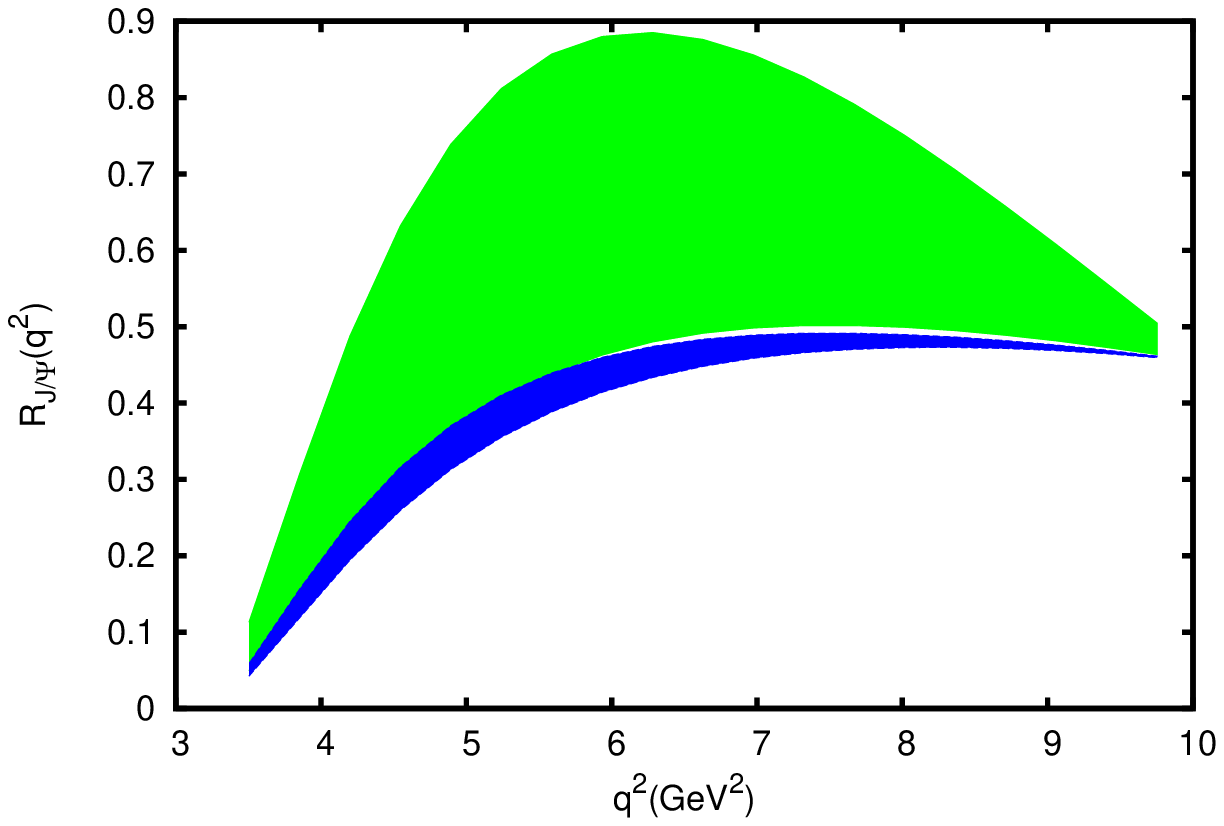}
\includegraphics[width=5cm,height=4cm]{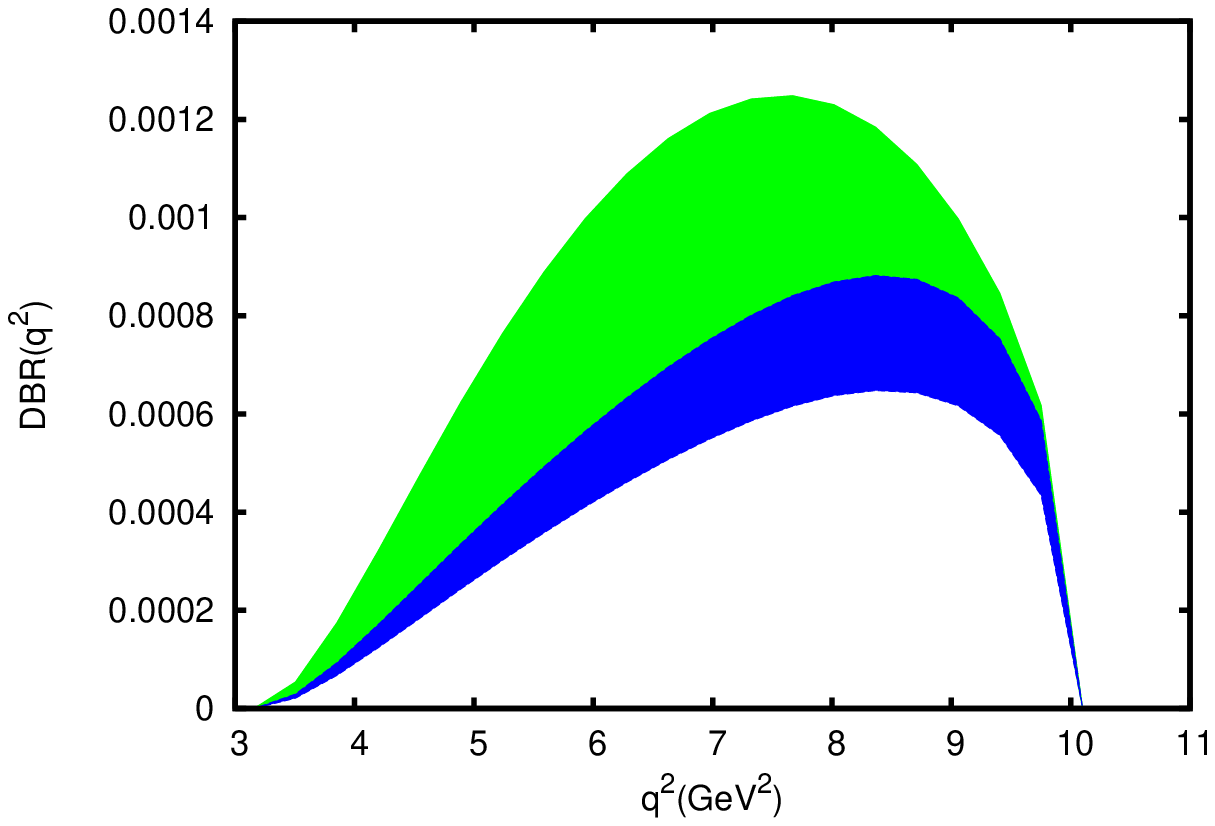}
\includegraphics[width=5cm,height=4cm]{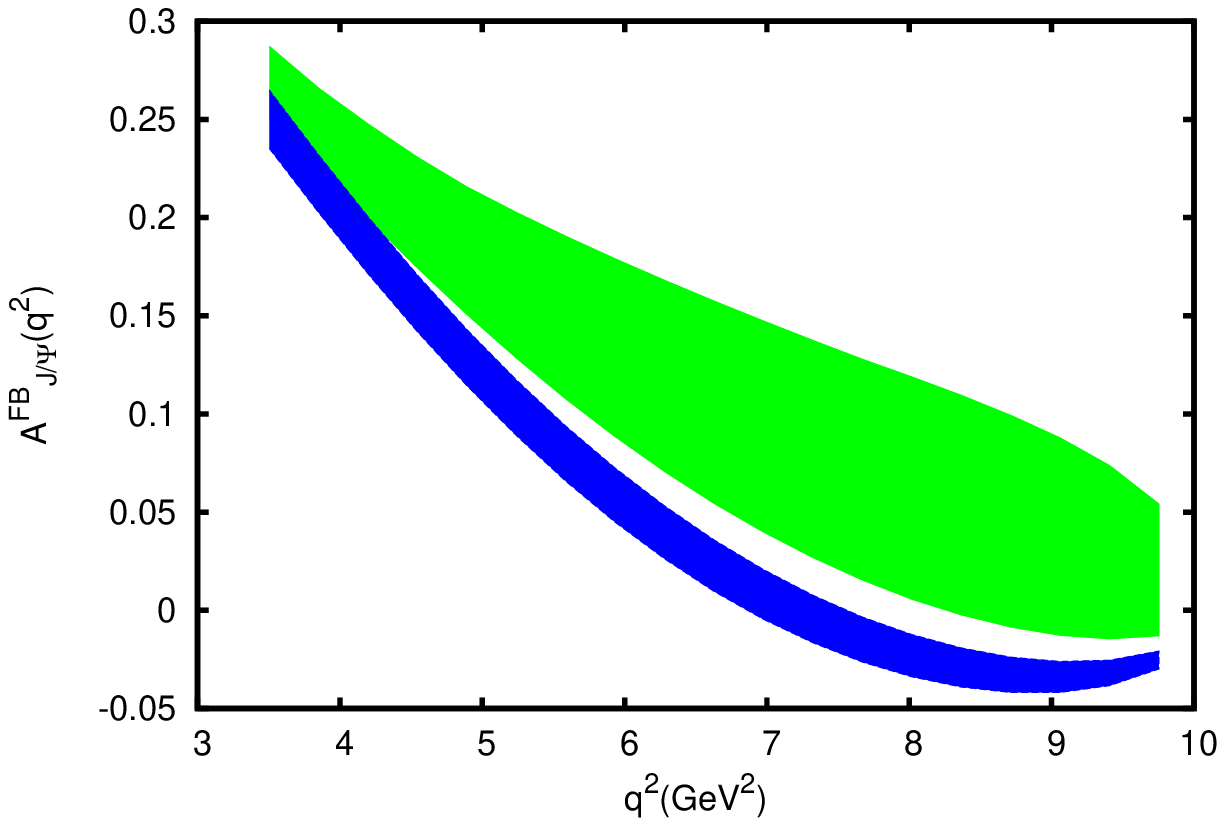}
\end{center}
\caption{Range in various $q^2$ dependent observables such as DBR$(q^2)$, $R(q^2)$, and $A^{FB}(q^2)$ for the $B_c \to \eta_c\tau\nu$~
(upper panel) and $B_c \to J/\Psi\tau\nu$~(lower panel) decays. The allowed range is each observable is shown in light~(green)
band once the NP couplings $(S_L,\,S_R)$ are varied within the allowed ranges shown in the left panel of Fig.~\ref{slsr}. We show 
in dark~(blue) band the corresponding SM prediction.}
\label{obs_slsr}
\end{figure}

\begin{table}[htdp]
\begin{center}
\begin{tabular}{|c|c||c|c||c|c|}
\hline
Observables &  Range &Observables & Range &Observables & Range\\[0.2cm]
\hline
\hline
$\mathcal B(B_c \to \tau\nu)\times 10^2$ &$[13.84, 248.94]$ &$R_{\eta_c}$ & $[0.213, 0.706]$ & $P_{\tau}^{J/\Psi}$ &$[-0.405, 0.117]$\\[0.2cm]
\hline
$\mathcal B(B_c \to \eta_c\,\tau\nu)\times 10^3$ &$[1.05, 3.02]$ &$R_{J/\Psi}$ &$[0.299, 0.486]$ & $P_{\tau}^D$ &$[0.301, 0.597]$\\[0.2cm]
\hline
$\mathcal B(B_c \to J/\Psi\,\tau\nu)\times 10^3$ &$[3.08, 5.71]$ &$P_{\tau}^{\eta_c}$ &$[0.053, 0.714]$ & $P_{\tau}^{D^{\ast}}$ & 
$[-0.090, -0.398]$\\[0.2cm]
\hline
\hline
\end{tabular}
\end{center}
\caption{Allowed ranges of various observables in the presence of $S_L$ and $S_R$ NP couplings}
\label{tab5}
\end{table}
Now we wish to see the effect of $S_L$ and $S_R$ NP couplings on various $q^2$ dependent observables such as ratio of branching 
ratio $R(q^2)$, forward backward asymmetry $A^{FB}(q^2)$, and the differential branching ratio ${\rm DBR}(q^2)$.  
The effect of NP couplings on these observables are shown in Fig.~\ref{obs_slsr}. Significant deviation from the SM expectation is
observed for all the observables in this scenario. We see that, in this scenario, all the observables are quite sensitive to the NP
couplings for $B_c \to \eta_c\tau\nu$ and $B_c \to J/\Psi\tau\nu$ decay modes. We also observe that, although, in the SM there is no 
zero crossing in the forward backward asymmetry parameter for the $B_c \to \eta_c\tau\nu$ decays; however, depending on the value of 
new scalar couplings $S_L$ and $S_R$, we might observe a zero crossing for this decay mode. 

\subsection{Scenario III: only $\widetilde{V}_L$ and $\widetilde{V}_R$ type NP couplings}
\label{vlt_vrt}
\begin{figure}
\begin{center}
\includegraphics[width=6cm,height=5cm]{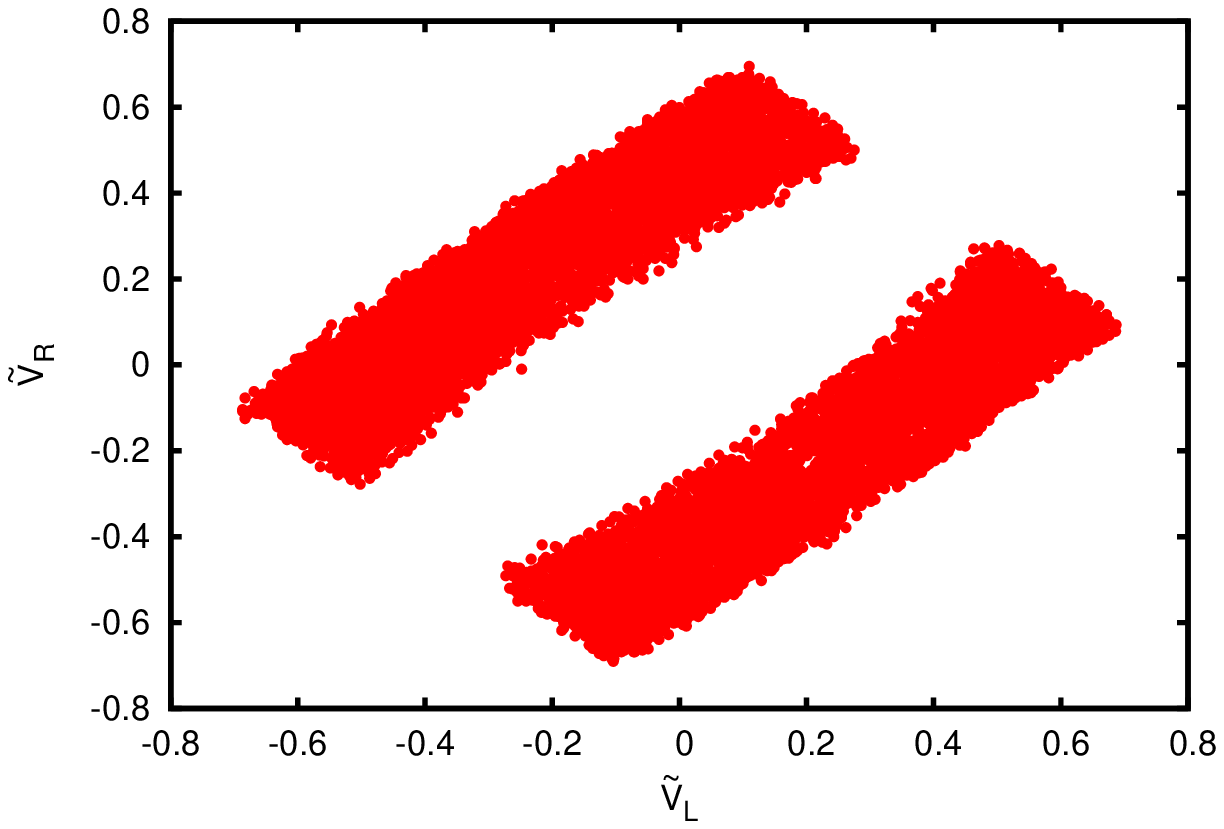}
\includegraphics[width=6cm,height=5cm]{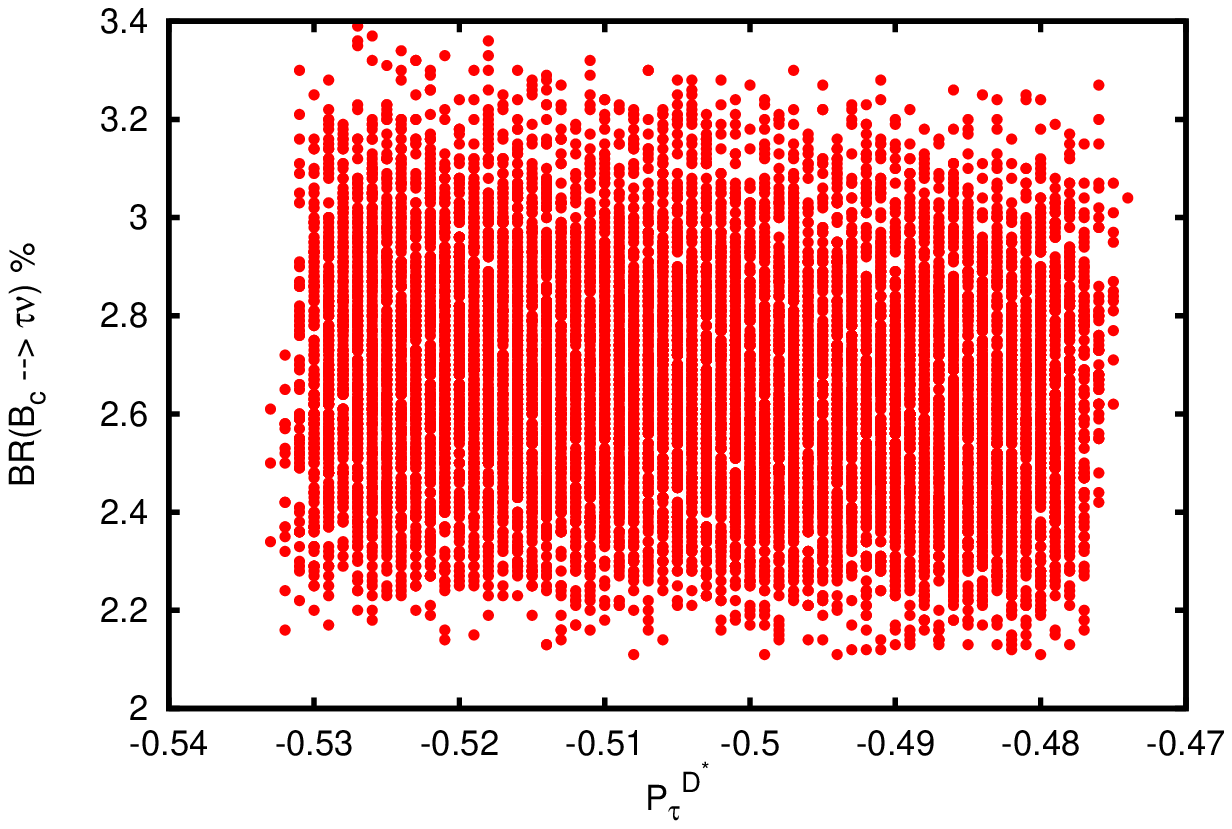}
\end{center}
\caption{Allowed ranges of $\widetilde{V}_L$ and $\widetilde{V}_R$ NP couplings are shown in the left panel once $2\sigma$ constraint 
coming from the measured
values of the ratio of branching ratios $R_D$ and
$R_{D^{\ast}}$ is imposed. We show in the right panel the allowed ranges in $\mathcal B(B_c \to \tau\nu)$ and $P_{\tau}^{D^{\ast}}$ in
the presence of these NP couplings.}
\label{vltvrt}
\end{figure}

In this scenario, we wish to see the effect of right handed neutrino couplings $\widetilde{V}_L$ and $\widetilde{V}_R$ on various 
observables. To realize this we vary only $\widetilde{V}_L$ and $\widetilde{V}_R$ and fix all other NP couplings to zero. The allowed
ranges of $\widetilde{V}_L$ and $\widetilde{V}_R$ obtained by using the $2\sigma$ constraint coming from the measured values of the
ratio of branching ratios $R_D$ and $R_{D^{\ast}}$ are shown in the left panel of Fig.~\ref{vltvrt}. The effect of 
$\widetilde{V}_L$ and $\widetilde{V}_R$ NP couplings on various observables are reported in Table.~\ref{tab6}. We also show, in 
particular, the effect of $\widetilde{V}_L$ and $\widetilde{V}_R$ on the branching ratio of $B_c \to \tau\nu$ and the on tau 
polarization fraction $P_{\tau}^{D^{\ast}}$ in the right panel of Fig.~\ref{vltvrt}. Although, very recently BELLE has reported their
results on $P_{\tau}^{D^{\ast}}$, the error is quite large. More precise data on $P_{\tau}^{D^{\ast}}$ in
future will help constraining the NP parameter space even more. Tau polarization fractions $P_{\tau}^D$ and $P_{\tau}^{\eta_c}$ do 
not vary at all with these NP couplings. It is expected since $B_c \to \eta_c\tau\nu$ and $B \to D\tau\nu$ decays depend only on 
$\widetilde{G}_V$ and hence the NP effect gets cancelled in the ratios. Deviation from the SM expectation observed in this scenario 
is quite similar to the deviations observed in scenario I of section.~\ref{vl_vr}.

\begin{table}[htdp]
\begin{center}
\begin{tabular}{|c|c||c|c||c|c|}
\hline
Observables &  Range &Observables & Range &Observables & Range\\[0.2cm]
\hline
\hline
$\mathcal B(B_c \to \tau\nu)\times 10^2$ &$[2.11, 3.39]$ &$R_{\eta_c}$ & $[0.238, 0.690]$ & $P_{\tau}^{J/\Psi}$ &$[-0.434, -0.492]$\\[0.2cm]
\hline
$\mathcal B(B_c \to \eta_c\,\tau\nu)\times 10^3$ &$[1.11, 3.07]$ &$R_{J/\Psi}$ &$[0.296, 0.416]$ & $P_{\tau}^D$ &$[0.334, 0.338]$\\[0.2cm]
\hline
$\mathcal B(B_c \to J/\Psi\,\tau\nu)\times 10^3$ &$[3.08, 5.19]$ &$P_{\tau}^{\eta_c}$ &$[0.141, 0.530]$ & $P_{\tau}^{D^{\ast}}$ & 
$[-0.474, -0.533]$\\[0.2cm]
\hline
\hline
\end{tabular}
\end{center}
\caption{Allowed ranges of various observables in the presence of $\tilde{V}_L$ and $\tilde{V}_R$ NP couplings}
\label{tab6}
\end{table}

\begin{figure}
\begin{center}
\includegraphics[width=5cm,height=4cm]{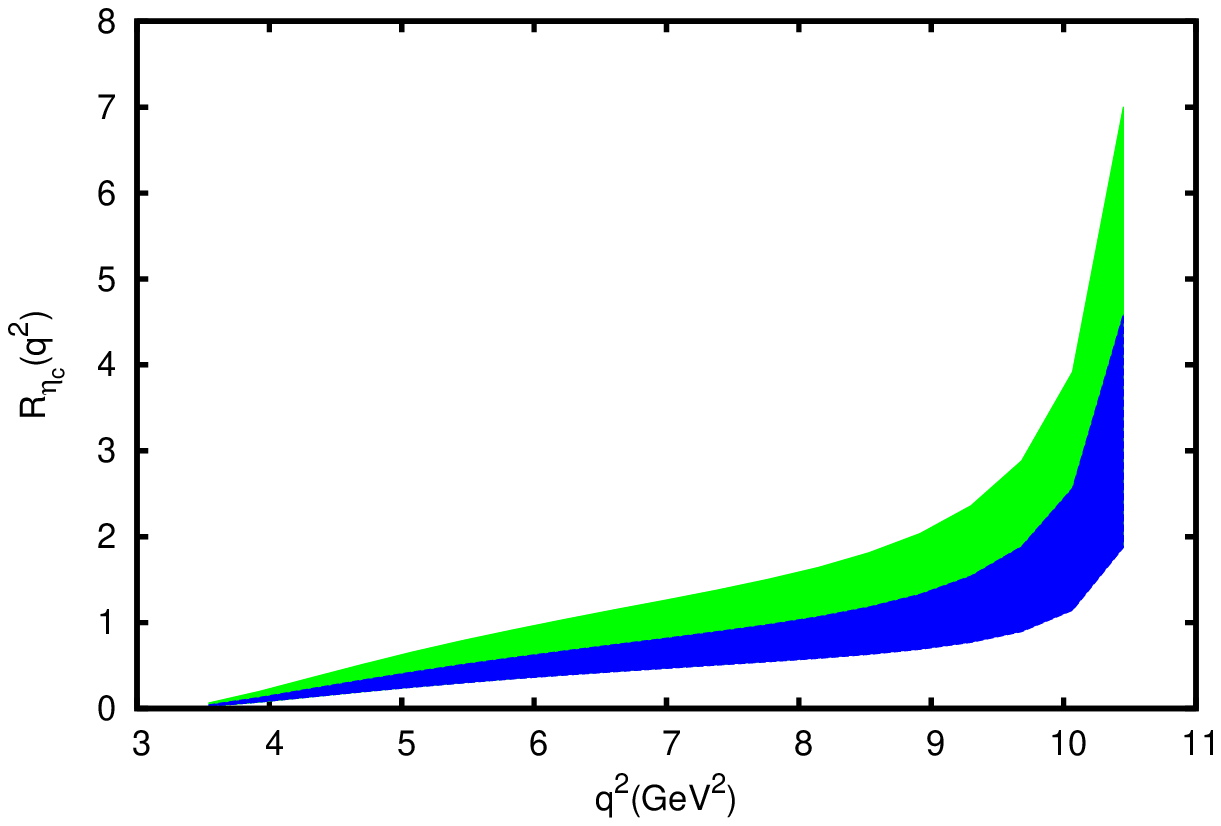}
\includegraphics[width=5cm,height=4cm]{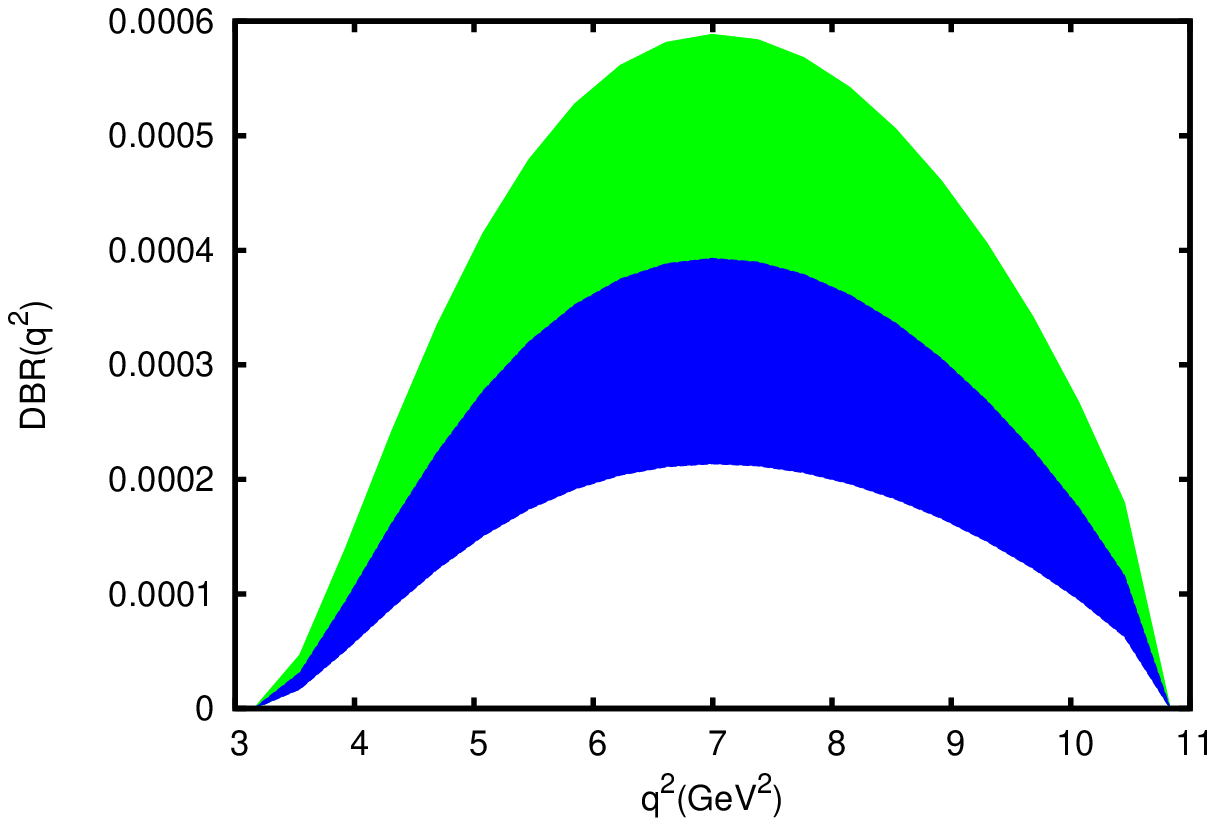}
\includegraphics[width=5cm,height=4cm]{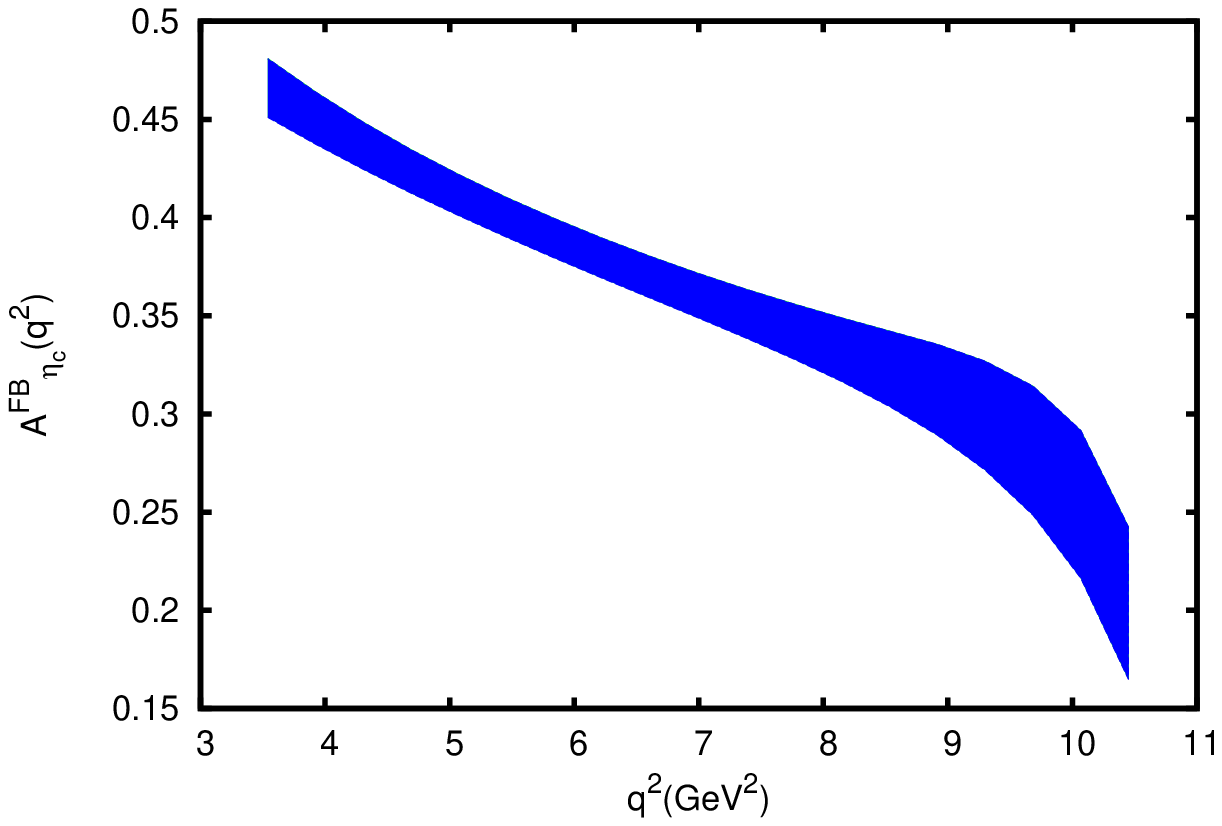}
\includegraphics[width=5cm,height=4cm]{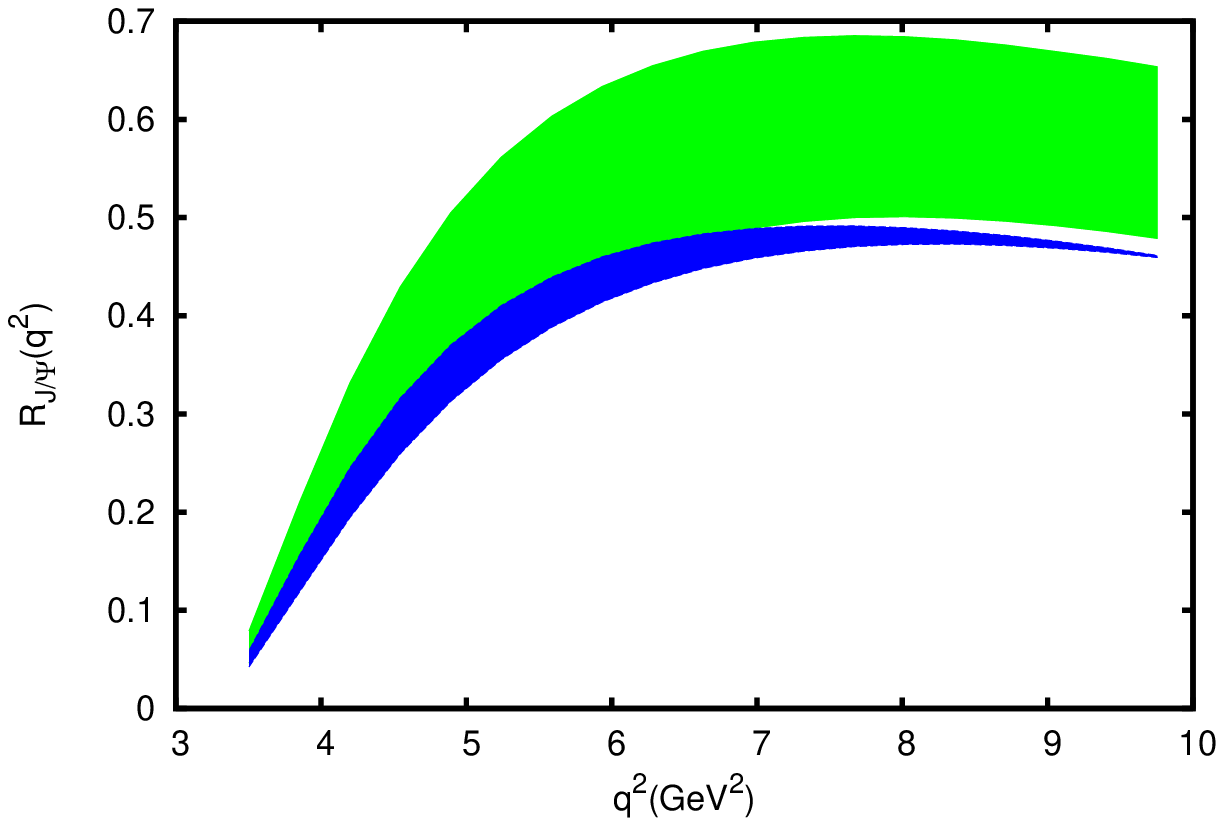}
\includegraphics[width=5cm,height=4cm]{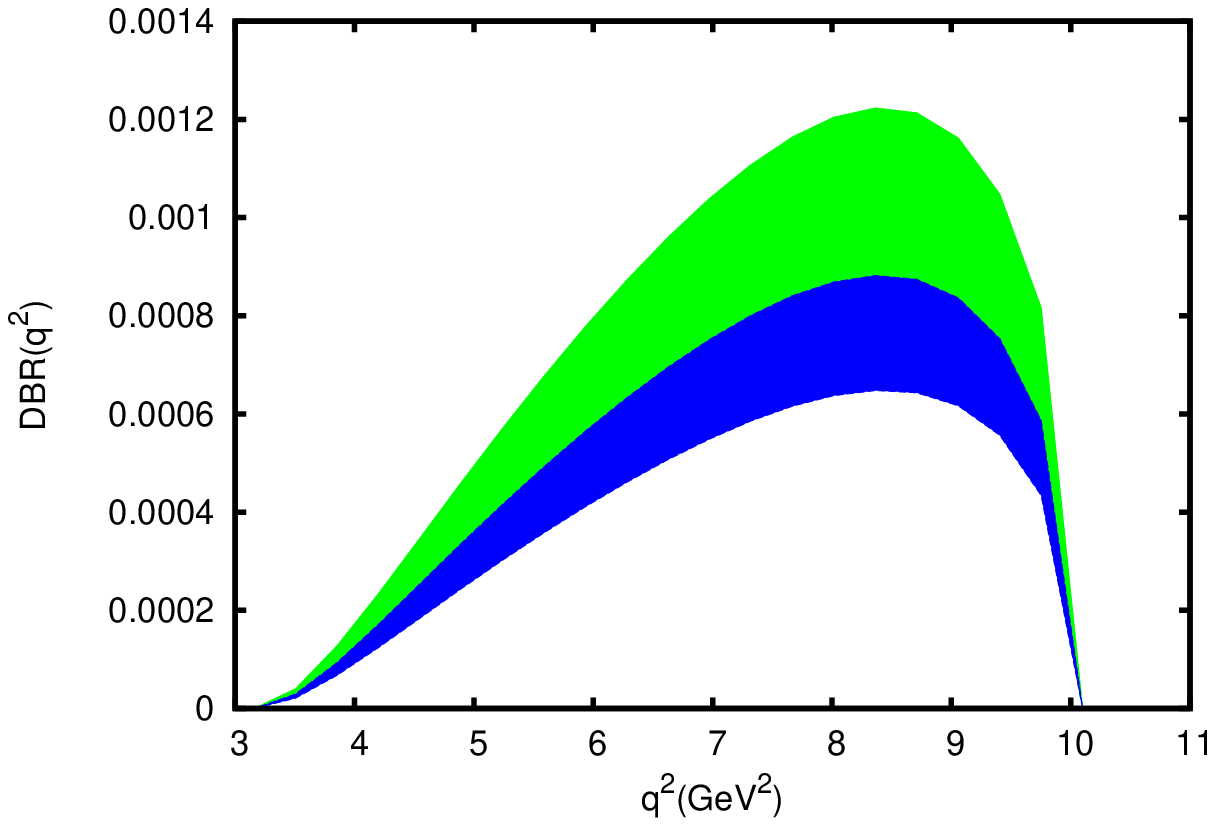}
\includegraphics[width=5cm,height=4cm]{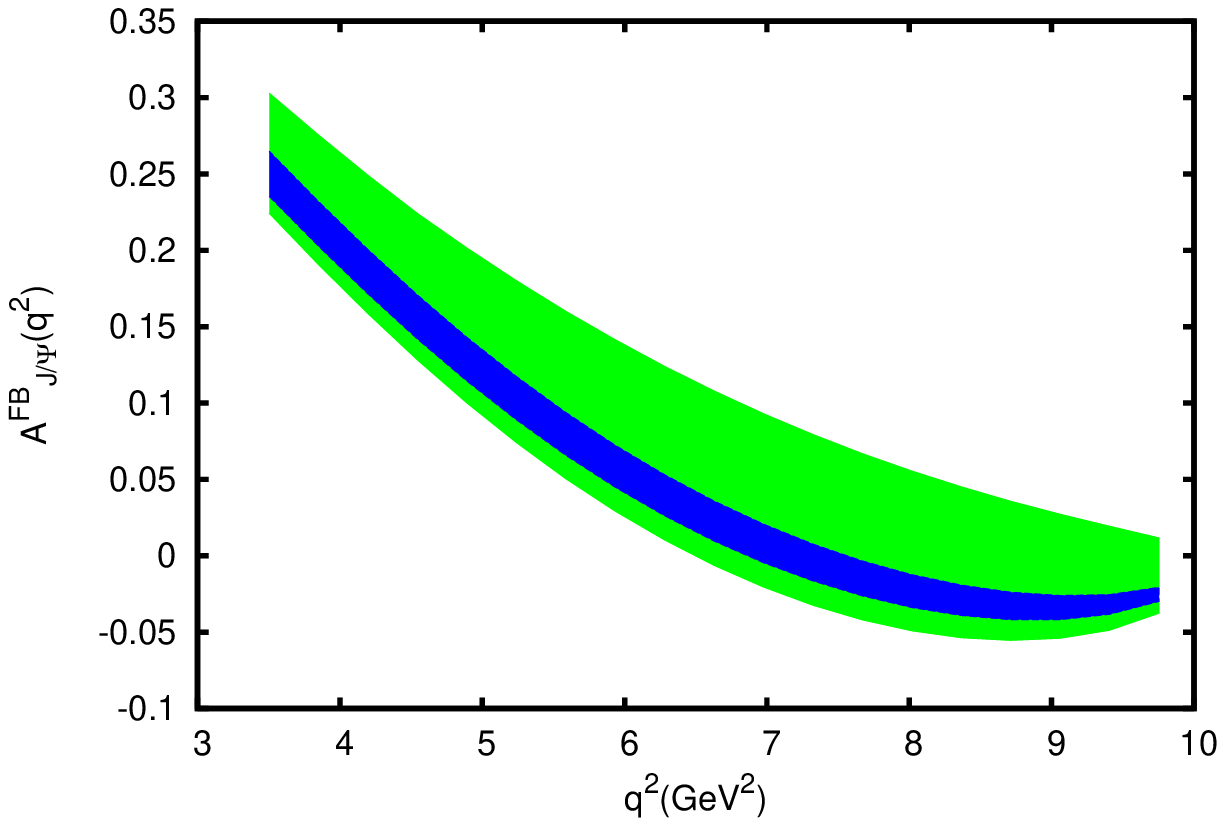}
\end{center}
\caption{Range in various $q^2$ dependent observables such as DBR$(q^2)$, $R(q^2)$, and $A^{FB}(q^2)$ for the $B_c \to \eta_c\tau\nu$~
(upper panel) and $B_c \to J/\Psi\tau\nu$~(lower panel) decays. The allowed range is each observable is shown in light~(green)
band once the NP couplings $(\widetilde{V}_L,\,\widetilde{V}_R)$ are varied within the allowed ranges shown in the left panel of 
Fig.~\ref{vltvrt}. We show in dark~(blue) band the corresponding SM prediction.}
\label{obs_vltvrt}
\end{figure}

The allowed ranges of various $q^2$ dependent observables such as ratio of branching ratio $R(q^2)$, the forward backward
asymmetry $A^{FB}(q^2)$, and the differential branching ratio ${\rm DBR}(q^2)$ are shown in Fig.~\ref{obs_vltvrt}. The SM
prediction is shown in dark~(blue) band whereas, the effect of NP couplings is shown in light~(green) band. The $q^2$ distribution 
looks quite similar
to what we obtain in scenario I of section~\ref{vl_vr}. Although we see a significant deviation of all the observables in this scenario, 
the forward
backward asymmetry parameter $A^{FB}_{\eta_c}(q^2)$ for the $B_c \to \eta_c\tau\nu$ decay mode does not seem to vary with the 
$\tilde{V}_L$ and $\tilde{V}_R$ NP couplings. This is obvious because the $B_c \to \eta_c\tau\nu$ differential branching ratio 
depends only on $\widetilde{G}_V$ and hence the NP effect gets cancelled in the ratio.  
On the other hand, $B_c \to J/\Psi\tau\nu$ decay differntial branching ratio depends not only 
on $\widetilde{G}_V$ but also on $\widetilde{G}_A$ and no such cancellation of the NP effects in the forward backward asymmetry parameter 
occurs for this decay mode. Hence we observe a significant deviation of $A^{FB}_{J/\Psi}$ from the SM expectation.

\subsection{Scenario IV: only $\widetilde{S}_L$ and $\widetilde{S}_R$ type NP couplings }
\begin{figure}
\begin{center}
\includegraphics[width=6cm,height=5cm]{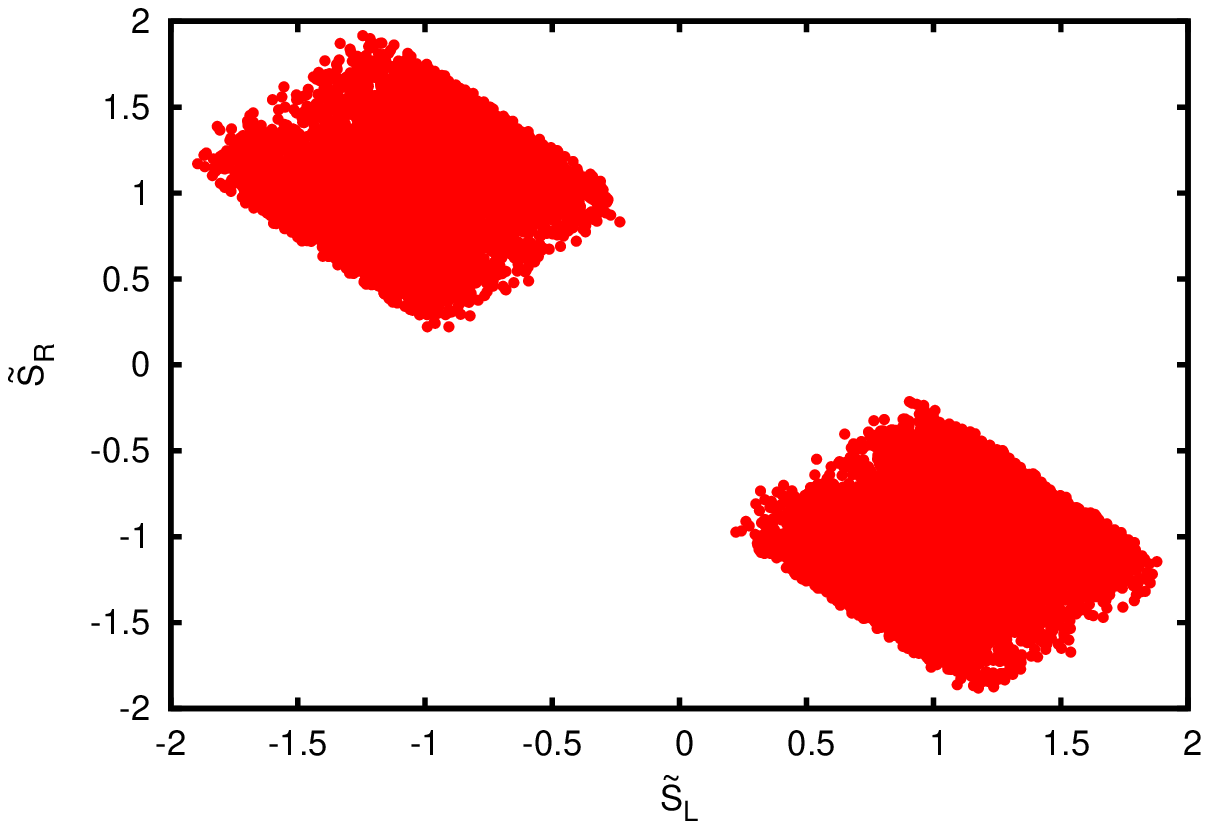}
\includegraphics[width=6cm,height=5cm]{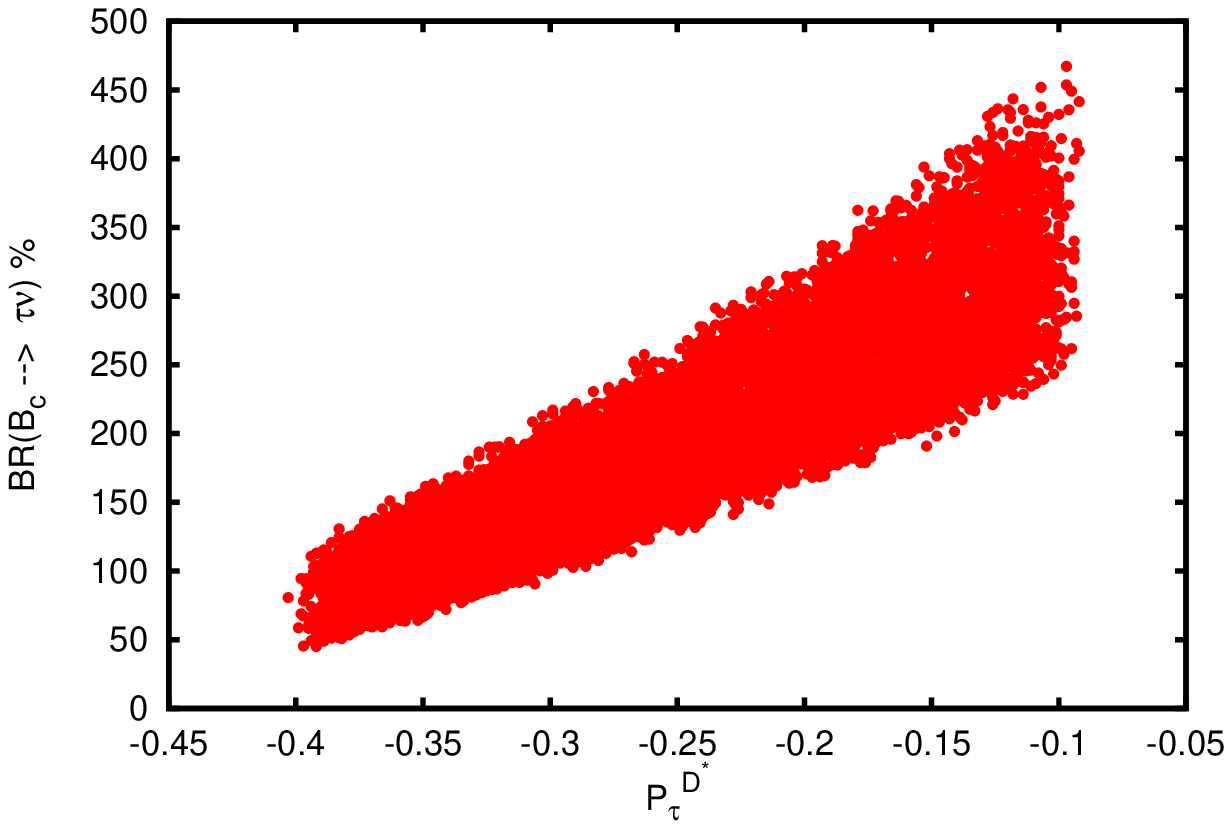}
\end{center}
\caption{Allowed ranges of $\widetilde{S}_L$ and $\widetilde{S}_R$ NP couplings are shown in the left panel once $2\sigma$ constraint 
coming from the measured
values of the ratio of branching ratios $R_D$ and
$R_{D^{\ast}}$ is imposed. We show in the right panel the allowed ranges in $\mathcal B(B_c \to \tau\nu)$ and $P_{\tau}^{D^{\ast}}$ in
the presence of these NP couplings.}
\label{sltsrt}
\end{figure}

To see the effect of new $\widetilde{S}_L$ and $\widetilde{S}_R$ couplings, associated with right handed neutrino, on various 
observables we vary $\widetilde{S}_L$ and $\widetilde{S}_R$ while keeping all other NP couplings to zero. We impose the $2\sigma$
constraint coming from the measured values of the ratio of branching ratios $R_D$ and $R_{D^{\ast}}$ and the resulting allowed ranges
of $\widetilde{S}_L$ and $\widetilde{S}_R$ NP couplings are shown in the left panel of Fig.~\ref{sltsrt}. The decay rate depends on 
$\widetilde{S}_L$ and $\widetilde{S}_R$ NP couplings quadratically and we obtain a less constrained NP parameter space. We also show
in the right panel of Fig.~\ref{sltsrt} the allowed ranges of $\mathcal B(B_c \to \tau\nu)$ and the tau polarization fraction 
$P_{\tau}^{D^{\ast}}$. The branching ratio of $B_c \to \tau\nu$ decays obtained in this scenario is rather large; more than $45\%$. 
However, from the total decay width of $B_c$ meson one can infer that branching ratio of $B_c \to \tau\nu$ decays should not be more
than $5\%$. Even if we relax the constraint upto $30\%$, the $\widetilde{S}_L$ and $\widetilde{S}_R$ NP couplings are ruled out 
although it can explain the anomalies persisted 
in the ratio of branching ratios $R_D$ and $R_{D^{\ast}}$. The allowed ranges of each observable obtained in this scenario are reported 
in Table.~\ref{tab7}. All the observables are very sensitive to the new $\widetilde{S}_L$ and $\widetilde{S}_R$ NP couplings.

\begin{table}[htdp]
\begin{center}
\begin{tabular}{|c|c||c|c||c|c|}
\hline
Observables &  Range &Observables & Range &Observables & Range\\[0.2cm]
\hline
\hline
$\mathcal B(B_c \to \tau\nu)\times 10^2$ &$[45.14, 467.22]$ &$R_{\eta_c}$ & $[0.238, 0.696]$ & $P_{\tau}^{J/\Psi}$ &$[-0.402, 0.122]$\\[0.2cm]
\hline
$\mathcal B(B_c \to \eta_c\,\tau\nu)\times 10^3$ &$[1.10, 3.15]$ &$R_{J/\Psi}$ &$[0.299, 0.490]$ & $P_{\tau}^D$ &$[0.335, 0.596]$\\[0.2cm]
\hline
$\mathcal B(B_c \to J/\Psi\,\tau\nu)\times 10^3$ &$[3.08, 5.79]$ &$P_{\tau}^{\eta_c}$ &$[0.150, 0.710]$& $P_{\tau}^{D^{\ast}}$ & 
$[-0.092, -0.403]$\\[0.2cm]
\hline
\hline
\end{tabular}
\end{center}
\caption{Allowed ranges of various observables in the presence of $\tilde{S}_L$ and $\tilde{S}_R$ NP couplings}
\label{tab7}
\end{table}

\begin{figure}
\begin{center}
\includegraphics[width=5cm,height=4cm]{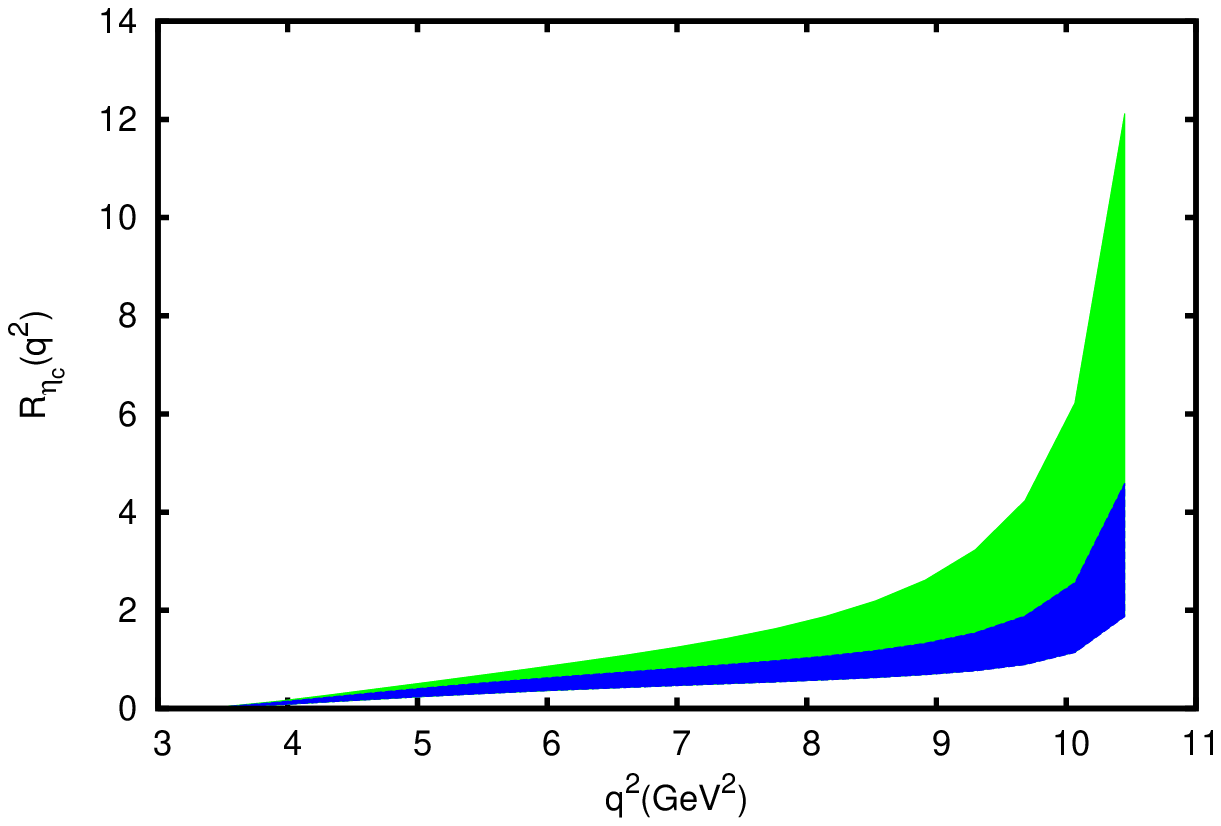}
\includegraphics[width=5cm,height=4cm]{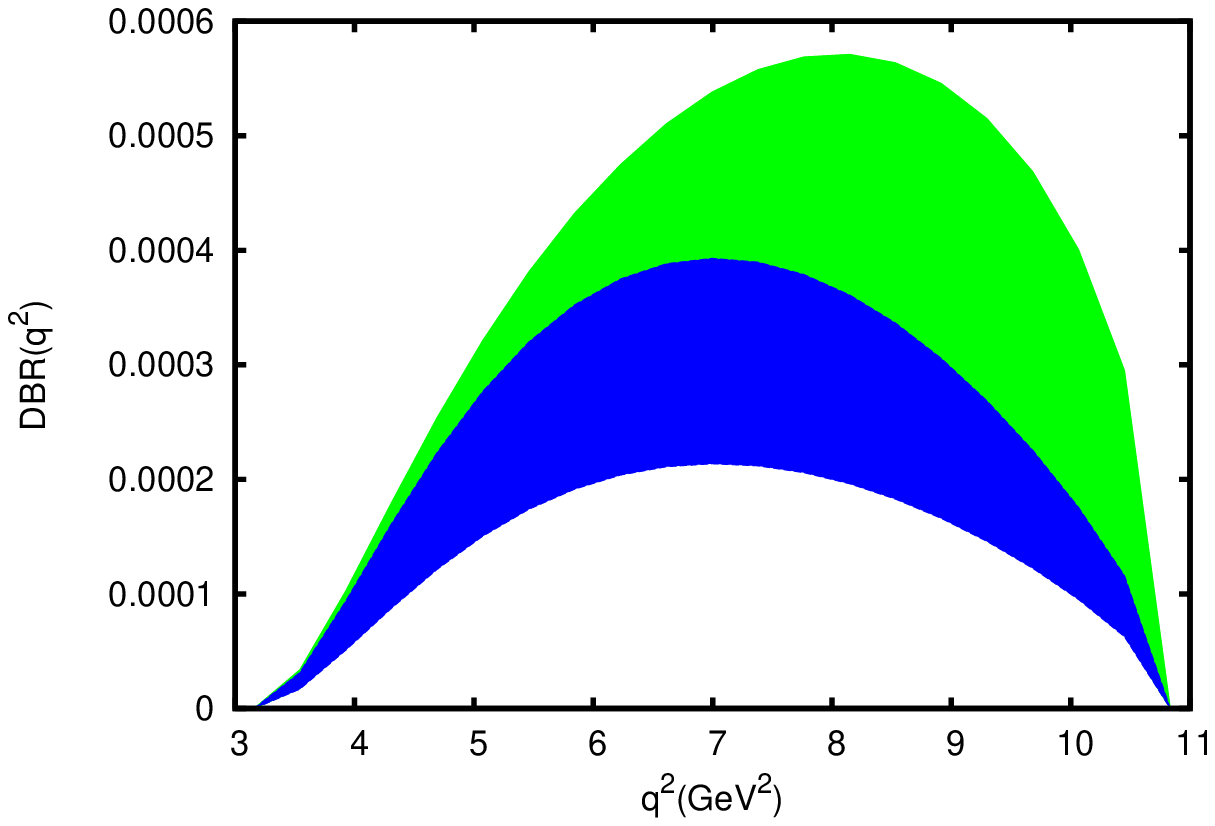}
\includegraphics[width=5cm,height=4cm]{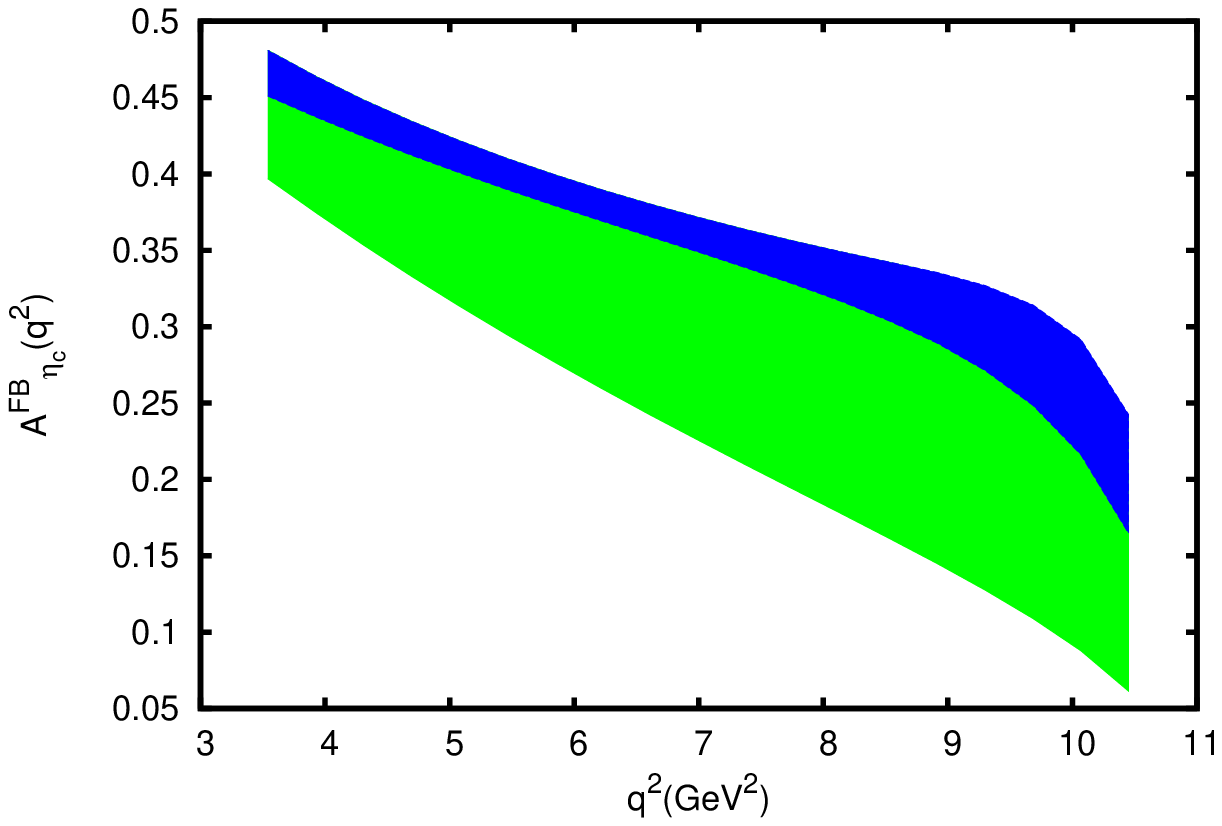}
\includegraphics[width=5cm,height=4cm]{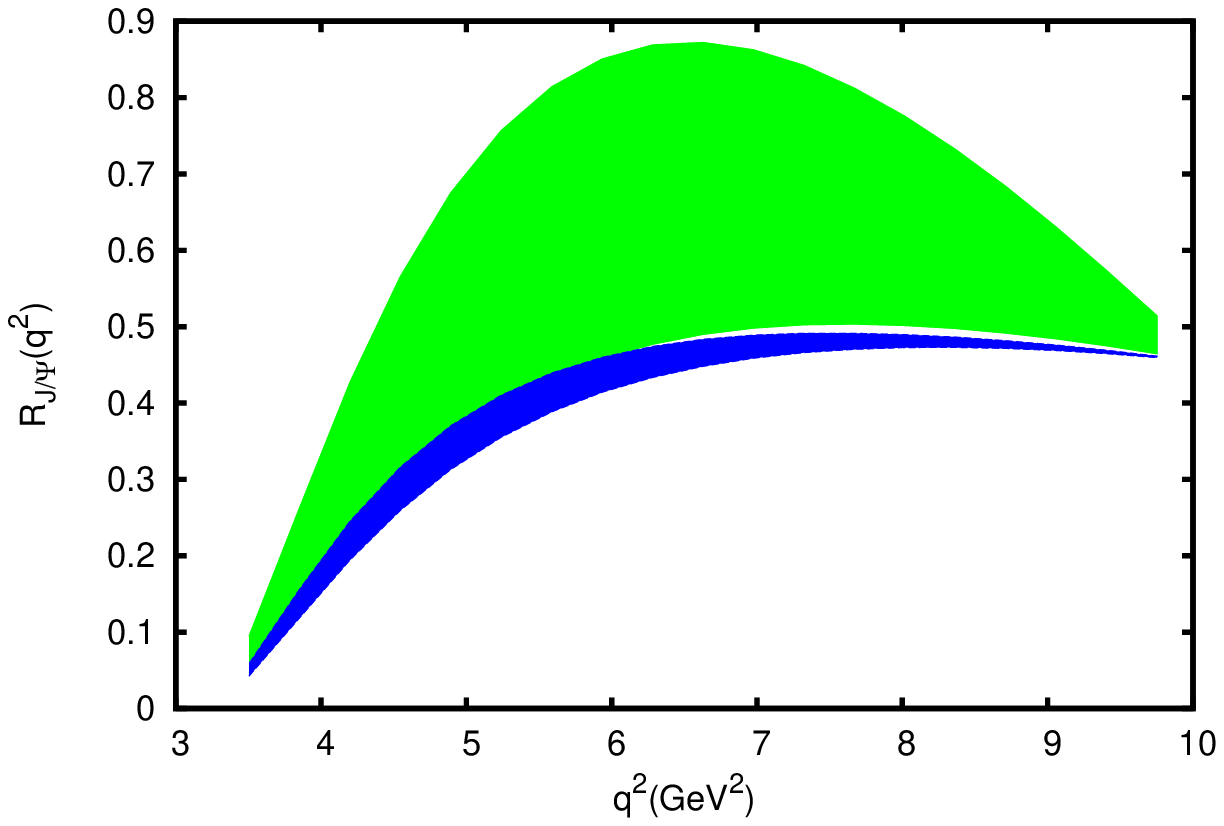}
\includegraphics[width=5cm,height=4cm]{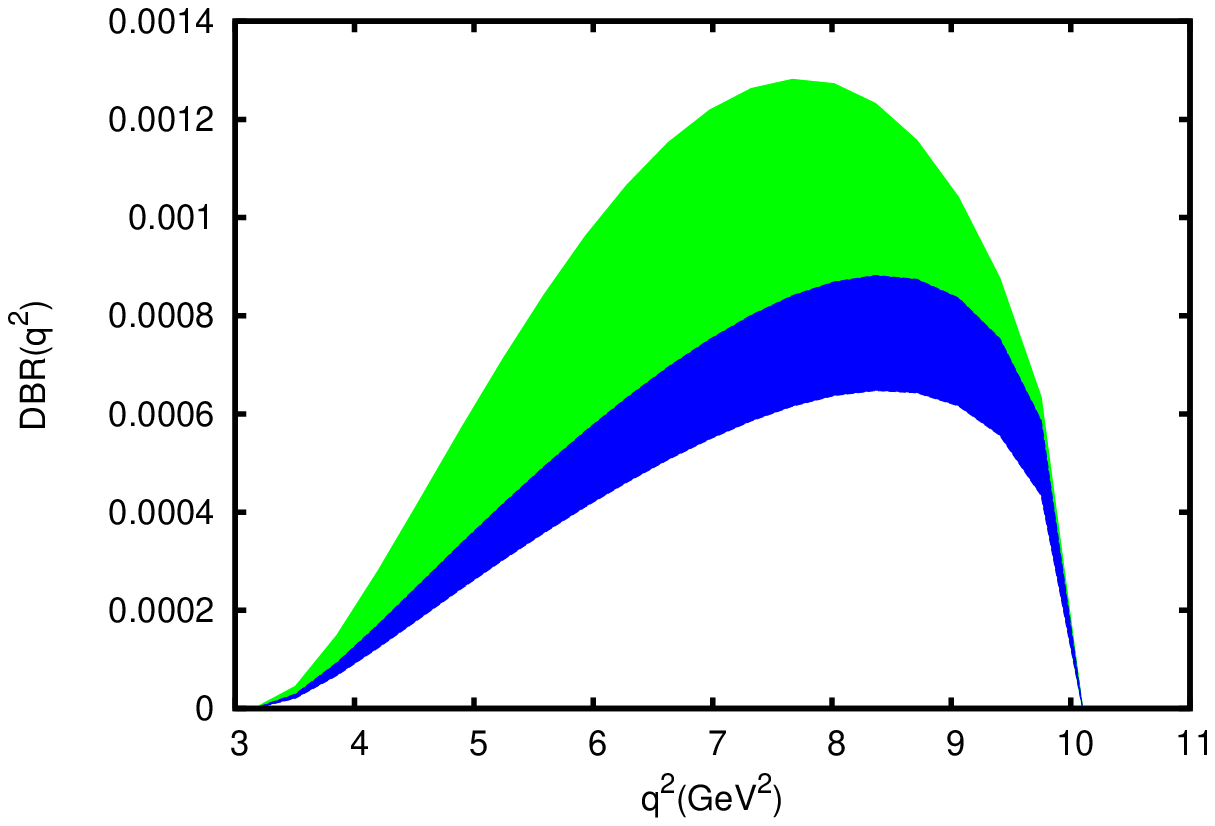}
\includegraphics[width=5cm,height=4cm]{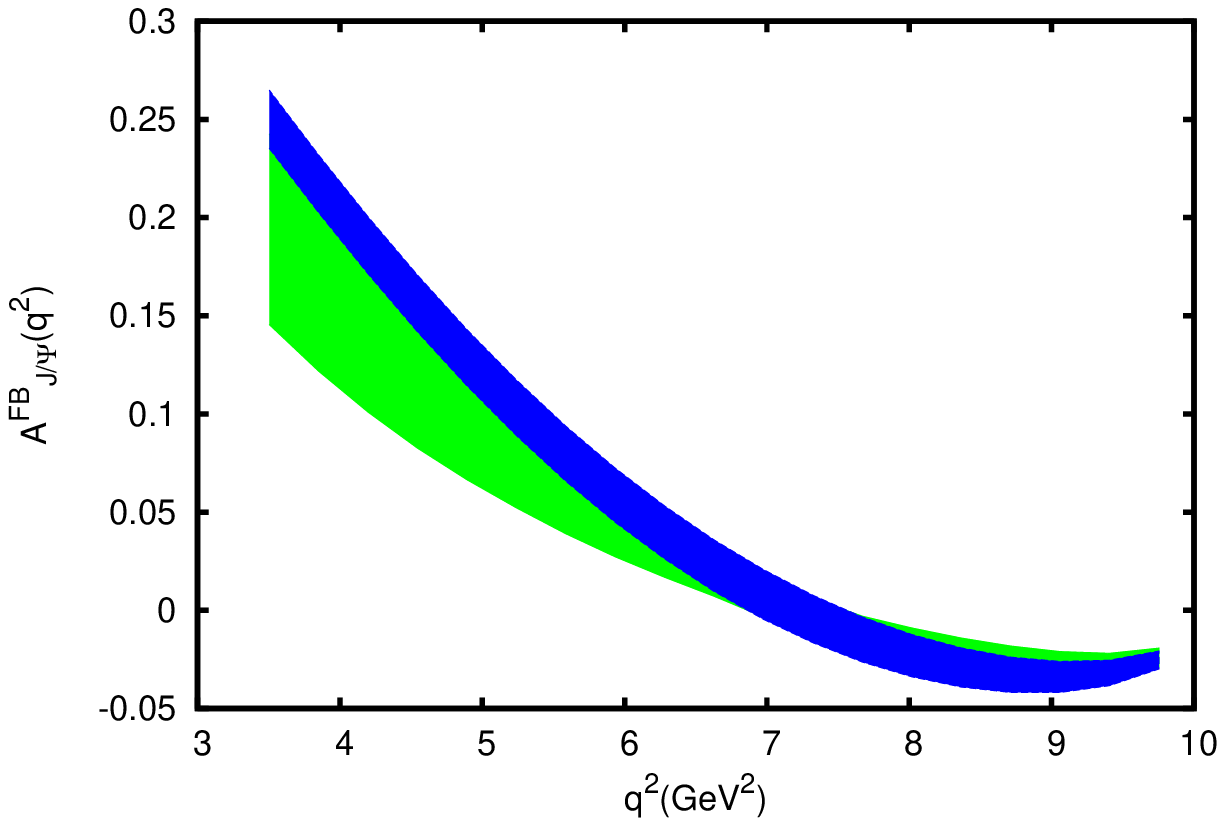}
\end{center}
\caption{Range in various $q^2$ dependent observables such as DBR$(q^2)$, $R(q^2)$, and $A^{FB}(q^2)$ for the $B_c \to \eta_c\tau\nu$~
(upper panel) and $B_c \to J/\Psi\tau\nu$~(lower panel) decays. The allowed range is each observable is shown in light~(green)
band once the NP couplings $(\widetilde{S}_L,\,\widetilde{S}_R)$ are varied within the allowed ranges shown in the left panel of 
Fig.~\ref{sltsrt}. We show in dark~(blue) band the corresponding SM prediction.}
\label{obs_sltsrt}
\end{figure}

We wish to see the effect of these NP couplings on various $q^2$ dependent observables for the $B_c \to \eta_c\tau\nu$ and 
$B_c \to J/\Psi\tau\nu$ decay modes. The allowed ranges of various observables such as $R(q^2)$, $A^{FB}(q^2)$, and ${\rm DBR}(q^2)$
are shown in Fig.~\ref{obs_sltsrt}. We see that all the observables deviate significantly from the SM expectation. Variation in 
$B_c \to \eta_c\tau\nu$ and $B_c \to J/\Psi\tau\nu$ decays, however, are quite different. This is what we expect because 
$B_c \to \eta_c\tau\nu$ 
decay branching ratio depends on these NP couplings through $\widetilde{G}_S$ term, whereas, $B_c \to J/\Psi\tau\nu$ decay branching
ratio depend on these NP couplings through $\widetilde{G}_P$ term. Although the effects of $\tilde{S}_L$ and $\tilde{S}_R$ NP couplings
are quite similar to $S_L$ and $S_R$ NP couplings of section.~\ref{sl_sr}, there are some differences. Unlike scenario II, we do not
observe any zero crossing in the $q^2$ distribution of the forward backward asymmetry parameter $A^{FB}_{\eta_c}$ in this scenario.

\section{Conclusion}
\label{con}
Deviations from the SM prediction have been observed not only in decays mediated via $b \to c$ charged current process but also 
in decays mediated via $b \to s$ neutral current process. In particular, the deviation of the measured ratios $R_D$ and $R_{D^{\ast}}$
from the SM prediction is more pronounced and it currently stands at $3.9\sigma$ level. Similarly, there are significant deviations from
the SM prediction in $b \to s\,l^+\l^-$ decays as well. The measured ratio $R_K$ deviates from the SM prediction by $2.6\sigma$. Again,
various other interesting tensions between the experimental results and SM prediction have been observed in rare $B \to K^{\ast}\mu^+\mu^-$
and $B \to \phi\mu^+\mu^-$ decays.
If it persists and confirmed by future experiments, these could provide the necessary information to unravel the
flavor structure of beyond the SM physics. Study of $B_c \to \eta_c\tau\nu$ and $B_c \to J/\Psi\tau\nu$ decays is interesting because
similar to $B \to (D,\,D^{\ast})\tau\nu$ decays, these decays are also mediated via $b \to c$ charged current interactions. Thus if NP is
present in $B \to (D,\,D^{\ast})\tau\nu$ decays, then it would show up in these decay modes as well. A detailed study of these decay modes
theoretically as well as experimentally is necessary in order to explore physics beyond the SM. Although, SM prediction of various
observables related to these decay modes has been reported by various authors, NP contribution has not been studied in details. To see
the effect of NP on various observables, we consider the most general effective Lagrangian in the presence of NP for the $b \to c\,l\,\nu$
process. We assume that NP is present only for the third generation leptons. We study four different NP scenarios. We summarise our 
results below. 

We first report the central values and the $1\sigma$ ranges of all the observables within the SM. The branching ratios of 
$B_c \to \eta_c\tau\nu$ and $B_c \to J/\Psi\tau\nu$ decays are at the order of $10^{-3}$. Again, we find the branching ratio of 
$B_c \to \tau\nu$ to be of the order of $2\%$. The values of ratio of branching ratios $R_{\eta_c}$ and $R_{J/\Psi}$ are quite similar
to the values reported in Ref.~\cite{Wen-Fei:2013uea}. We also give the first prediction of the tau polarization fraction $P_{\tau}^{\eta_c}$
and $P_{\tau}^{J/\Psi}$ for the $B_c \to \eta_c\tau\nu$ and $B_c \to J/\Psi\tau\nu$ decay modes.

We include vector-and scalar type NP interactions that involve both right handed as well as left handed neutrinos in our analysis and explore
four different NP scenarios. In the first scenario, we consider only vector type NP interactions that involve left handed neutrinos. We vary
$V_L$ and $V_R$ while keeping all other NP couplings to zero. Deviation from the SM expectation is observed for all the observables. 
The central value of $P_{\tau}^{D^{\ast}}$ reported by BELLE lies outside the allowed range of $P_{\tau}^{D^{\ast}}$ obtained in
this scenario. However, the uncertainty associated with the measured value of $P_{\tau}^{D^{\ast}}$ is rather large. More precise data 
in future on $P_{\tau}^{D^{\ast}}$ will
definitely help constraining the NP parameter space even more.
The allowed range of $\mathcal B(B_c \to \tau\nu)$ is consistent with the total decay width of $B_c$ meson.
We see no deviation from the SM prediction of tau polarization fraction $P_{\tau}^D$ and $P_{\tau}^{\eta_c}$ as the NP effects coming from
$V_L$ and $V_R$ couplings cancel in the ratios. We also see the effect of these NP couplings on various $q^2$ dependent observables. 
Significant deviation from the SM expectation is observed once the NP couplings are included. There is, however, no deviation from the
SM prediction of the forward backward asymmetry parameter $A^{FB}_{\eta_c}$.

In the second scenario, we consider that NP effect is due to the scalar type interactions that involves left handed neutrinos only, i.e,
$S_{L,\,R} \ne 0$, whereas all other NP couplings are zero. Significant deviation from the SM expectation is observed for all the 
observables. It is also worth mentioning that the tau
polarization $P_{\tau}^{D^{\ast}}$ deviates significantly from the central value reported by BELLE. Again, we notice that, in this 
scenario, for some particular values of $S_L$ and $S_R$, the value of
$\mathcal B(B_c \to \tau\nu)$ exceeds the total decay width of $B_c$ meson. However, only less than 
$5\%$ of the total decay width of $B_c$ meson can be explained by semi(taunic) mode. Even if we relaxed the constraint upto $30\%$, 
a substantial part of NP parameter space can be excluded. Hence, $B_c$ total decay width put a severe constraint on $S_L$ and $S_R$ type
NP couplings. We also see the effect of NP couplings on various $q^2$ dependent observables. The deviation observed in this scenario is
more pronounced than the deviation observed in scenario I.

In the third scenario, we set $\widetilde{V}_{L,\,R} \ne 0$ while keeping all other NP couplings to zero. Similar to scenario I, we see
significant deviation of all the observables from the SM prediction. We want to mention that, branching ratio of $B_c \to \tau\nu$ 
obtained in this scenario is consistent with the experimentally measured total decay width of $B_c$ meson. Again, although the central
value of $P_{\tau}^{D^{\ast}}$ reported by BELLE lies outside the allowed range obtained, however, the $1\sigma$ range of the experimental
value does overlap with the allowed range. More precise data on $P_{\tau}^{D^{\ast}}$ observable is needed to constrain the NP parameter
even further. The deviation in various $q^2$ dependent observables observed in this scenario is similar to the ones that we observed in 
scenario I. The forward backward asymmetry parameter $A^{FB}_{\eta_c}$ does not vary at all as the NP dependency cancels in the ratio.

In the fourth scenario we consider only $\widetilde{S}_L$ and $\widetilde{S}_R$ type NP couplings. Again, as expected, the deviations
from the SM prediction in this scenario is quite high. 
We notice that the branching ratio of $B_c \to \tau\nu$ decays obtained in this scenario is rather large; more than $45\%$. 
However, from the total decay width of $B_c$ meson one can infer that branching ratio of $B_c \to \tau\nu$ decays should not be more
than $5\%$. Even if the constraint is relaxed upto $30\%$, the $\widetilde{S}_L$ and $\widetilde{S}_R$ NP couplings are ruled out although 
it can explain the anomalies persisted 
in the ratio of branching ratios $R_D$ and $R_{D^{\ast}}$. It is worth mentioning that, 
all the observables are very sensitive to the new $\widetilde{S}_L$ and $\widetilde{S}_R$ NP couplings, similar to scenario II. 
All the $q^2$ dependent observables are also very sensitive to the new  $\widetilde{S}_L$ and $\widetilde{S}_R$ NP couplings.

In conclusion, we observe that, $B_c$ lifetime put a severe constraint on $S_{L,\,R}$ and $\widetilde{S}_{L,\,R}$ type NP couplings. 
More precise calculations of the $B_c$ lifetime and measurements of the branching
fractions of its various decay channels in future should help constrain the NP parameter space even further.
Again, the observable 
$P_{\tau}^{D^{\ast}}$ has the potential to distinguish between various NP scenarios once more precise data is available. At present,
however, the experimental uncertainty associated with the tau polarization fraction $P_{\tau}^{D^{\ast}}$ is rather large. More
precise data in future will difinitely help identifying the nature of NP. Measurement of all the observables for the 
$B_c \to \eta_c\tau\nu$ and $B_c \to J/\Psi\tau\nu$ decay modes will be crucial to explore the nature of NP patterns.

\end{document}